\documentclass[12pt,preprint]{aastex6}

\usepackage{epsfig}

\def\fluxthres{\hat f_{\bar \e}}
\def\fluxeps{f_{_{\rm \epsilon}}}

\def\zmax{z_{\rm max}}

\def\e{\epsilon}

\def\Estarg{{\cal E}_{*\gamma}}

\def\Swift{\emph{Swift}}
\def\Fermi{\emph{Fermi}}

\newcommand{\begeq}{\begin{equation}}
\newcommand{\fineq}{\end{equation}}
\newcommand{\begfig}{\begin{figure}}
\newcommand{\finfig}{\end{figure}}
\newcommand{\begeqarray}{\begin{eqnarray}}
\newcommand{\fineqarray}{\end{eqnarray}}

\slugcomment{accepted for publication in ApJ}

\shorttitle{Redshift Distribution of GRBs} 

\shortauthors{Le \& Mehta}

\begin{document}

\title{Revisiting the Redshift Distribution of Gamma-ray Bursts in the \Swift~Era}

\author{Truong Le and Vedant Mehta}

\affil{Department of Physics, Astronomy \& Geology, Berry College, Mount Berry, GA 30149, USA; tle@berry.edu}

\begin{abstract}

Le \& Dermer developed a gamma-ray burst (GRB) model to fit the redshift and the jet opening angle distributions measured with pre-\Swift~and \Swift~missions and showed that GRBs do not follow the star formation rate. Their fitted results were obtained without the opening angle distribution from \Swift~with an incomplete \Swift~sample, and that the calculated jet opening angle distribution was obtained by assuming a flat $\nu F_\nu$ spectrum.  In this paper, we revisit the work done by Le \& Dermer with an assumed  broken power law GRB spectrum. Utilizing more than 100 GRBs in the \Swift~sample that include both the observed estimated redshifts and jet opening angles, we obtain a GRB burst rate functional form that gives acceptable fits to the pre-\Swift~and \Swift~redshift and jet opening angle distributions with an indication that an excess of GRBs exists at low redshift below $z \approx 2$. The mean redshifts and jet opening angles for pre-\Swift~(\Swift) are $\langle z \rangle \sim 1.3$ ($1.7$) and $\langle \theta_{\rm j} \rangle \sim 7^o$ ($11^o$), respectively. Assuming a GRB rate density (SFR9), similar to the Hopkins \& Beacom star formation history and as extended by Li, the fraction of high-redshift GRBs is estimated to be below 10\% and 5\% at $z \geq 4$ and $z\geq 5$, respectively, and below 10\% at $z \leq 1$.

\end{abstract}


\section{Introduction}

Gamma-ray bursts (GRBs) are short and intense irregular pulses of gamma-ray radiation that last less than a few minutes with a nonthermal spectrum (broken power law) peaking at $\sim 10-10^4$ keV \citep[e.g.,][]{pre00}. Isotropically, GRBs radiate between $10^{48}$ and $10^{55}$ erg, but the true amount of energy released in these explosions is about $10^{48} - 10^{52}$ erg \citep[see][and references therein]{kz15}. The emissions of GRBs are believed to result from the jets \citep[e.g.,][]{spj99, sta99,bgp15}, but, the formation of jets in these objects is still unknown.

A GRB duration has two distinct peaks, one at 0.3 s and the other at about 30 s. Bursts with duration less than $T_{90} < 2$ s are classified as short GRBs (SGRBs), and those that last for more than $T_{90} > 2$ s are called long GRBs (LGRBs), where $T_{90}$ is the time interval of gamma-ray photons collected  (from $5\%$ to $95\%$ of the total GRB counts) by a given instrument \citep[e.g.,][]{kou93}. Based on the peak duration distribution, it was suspected that these peaks correspond to two physically distinct progenitors. That is, LGRBs result from the collapse of massive stars (mass $\gtrsim 15 \, M_\odot$), and  SGRBs result from mergers of two neutron stars or a neutron star and a black hole \citep[e.g.,][]{woo93}. However, the connection between the GRB classifications based on burst duration and based on distinct physical origins is still not fully understood \citep[see][for a recent review]{kz15}.

Since LGRBs are associated with the deaths of massive stars, it is commonly assumed that the GRB rate density follows the star formation rate (SFR) density history \citep[e.g.,][]{pac98,wp10}. Because of their high luminosity, up to $10^{54}$ erg s$^{-1}$ \citep[see][and references therein]{pes16}, GRBs can be detected out to the early universe \citep[e.g.,][]{lr00,bl02,bl06}, and the farthest GRB to date is GRB 090429B with a photometric redshift $z = 9.4$ \citep[][]{cuc11}. Hence, this holds promise that GRBs can be used to probe the early universe.

One of the most important properties to understand the central engines of GRBs is to know the energy budgets of these enormous explosions. One particular physical property of GRBs that has a large impact on the observed energies is the jet opening angle to which the jetted outflow is collimated. This angle requires observations of the achromatic jet break time in the power-law decay of the afterglow emission and the particle density profile of the surrounding circumburst medium of the host environment \citep[e.g.,][]{wax97,spj99}.  Unfortunately, the achromatic jet break time and the particle density profile of the surrounding circumburst medium are difficult to estimate \citep[e.g.,][]{kz15}. 

\citet{ld07} examined whether the differences between the pre-\Swift~and \Swift~redshift and jet opening angle distributions can be explained with a physical model for GRBs that takes into account the different flux thresholds of GRB detectors. In that model, they parameterized the jet opening angle distribution for an assumed flat $\nu F_\nu$ spectrum and found best-fit values for the $\gamma$-ray energy release for different functional forms of the comoving rate density of GRBs, assuming that the properties of GRBs do not change with time. Adopting the uniform jet model, they assumed that the energy per solid angle is roughly constant within a well-defined jet opening angle. Their results showed that an intrinsic distribution in the jet opening angles yielded a good fit to the pre-\Swift~and \Swift~redshift samples and provided an acceptable fit to the distribution of the opening angles measured with pre-\Swift~GRB detectors. In their work, they could not confirm the validity of the \Swift's observed opening angle distribution because limited observed jet opening angles were available. Their analysis showed that a good fit was only possible, however, by modifying the \citet{hb06} SFR to provide positive evolution of the SFR history of GRBs to high redshifts (e.g., SFR5 and SFR6; see Figure~\ref{fig1}a). The best-fitted values were obtained with the average beaming-corrected $\gamma$-ray energy release $\Estarg = 4 \times 10^{51}$ erg, the minimum and maximum jet opening angles $\theta_{j, min} = 0.05$ rad and $\theta_{j, max} = 0.7$ rad, and the jet opening angle power-law indices $s \approx -1.25$ or $-1.2$. The conclusion of their work suggested that GRB activity was greater in the past and is not simply proportional to the bulk of the star formation as traced by the blue and UV luminosity density of the universe.
\begfig[t] \hskip-0.25in \epsscale{1.15} \plottwo{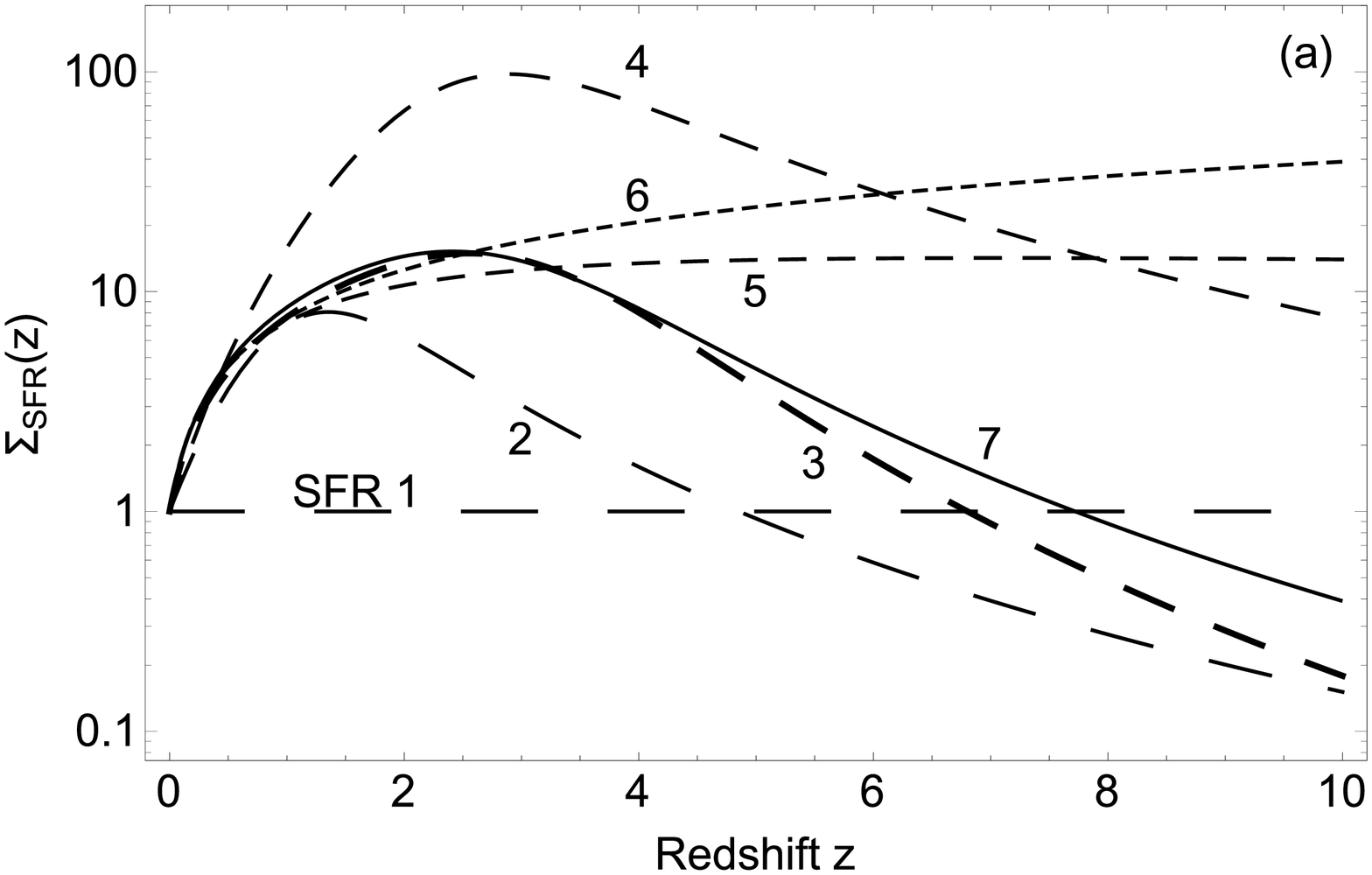}{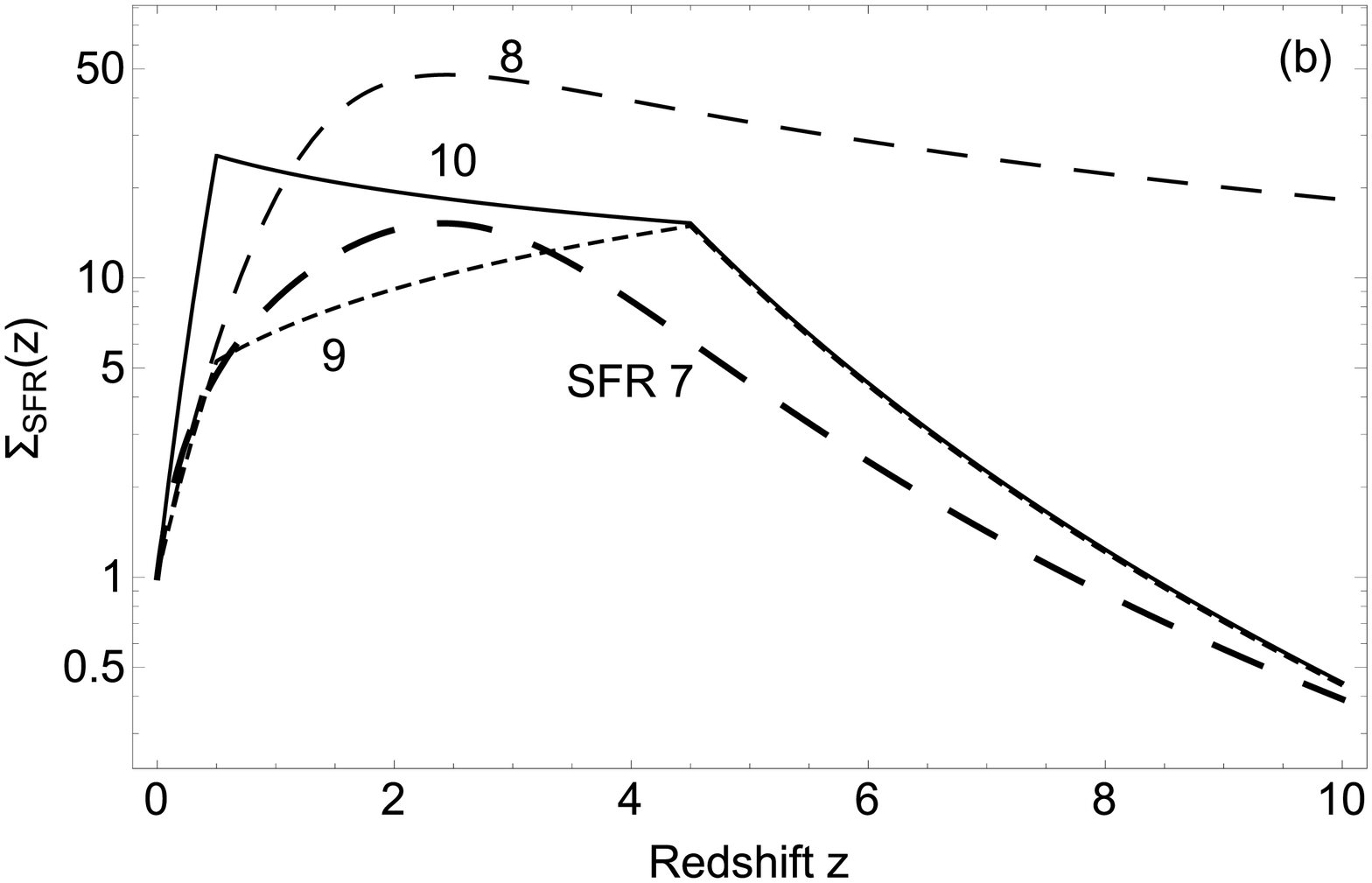}
\caption{\footnotesize (a) Star formation rate (SFR) function for GRBs, assumed to
be proportional to comoving SFR histories as shown. The long dashed line
(SFR1) is a constant comoving density. SFR2 and SFR4 are the lower
and upper SFRs given by Equation (\ref{eq13}) in~\citet{ld07}. SFR3 is
the~\citet{hb06} SFR history given by Equation (\ref{eq16}), with
($a_1=0.015$, $a_2=0.1$, $a_3=3.4$, and $a_4=5.5$). SFR5 ($a_1=0.015$,
$a_2=0.12$, $a_3=3.0$, $a_4=1.3$) and SFR6 ($a_1=0.011$, $a_2=0.12$,
$a_3=3.0$, $a_4=0.5$) are the GRB rates that give a good fit to the
\Swift~and pre-\Swift~redshift distribution assuming a GRB flat spectrum. SFR7 (using Equation~(\ref{eq16}) with $a_1=0.0157$, $a_2=0.118$, $a_3=3.23$, $a_4=4.66$) is the~\citet{hb06} SFR history and as extended by~\citet{li08}. (b) SFR8 (using Equation~(\ref{eq17}) with $a_1=0.005$, $a_2=4.5$, $a_3=1$) is a GRB rate required to fit the redshift and jet opening angle distributions for both pre-\Swift~and \Swift~incomplete samples assuming a flat GRB spectrum. SFR9 (for the incomplete sample) and SFR10 (for the complete sample) are the GRB density rates required to fit the redshift and jet opening angle distributions for both pre-\Swift~and \Swift~samples assuming a broken power law GRB spectrum with the low- and high-energy spectral indices $a = 1$ and $b = -0.5$, respectively. SFR9 and SFR10 use Equation~(\ref{eq18}) with $\alpha = 4.1, \beta = 0.8, \gamma = -5.1, z_1 = 0.5, z_2 = 4.5$ and $\alpha = 8, \beta = -0.4, \gamma = -5.1, z_1 = 0.5, z_2 = 4.5$, respectively.}
\label{fig1} 
\finfig

The result of the GRB density history at high redshift from \citet{ld07} was consistent with that of \citet{mes06}, \citet{dai06}, and \citet{gp07} and was later confirmed by many other researchers \citep[e.g.,][]{kis08,yuk08,kis09,wp10,vir11,jak12,wan13}. However, it is unclear whether this excess at high redshift is due to luminosity evolution \citep[][]{sc07,sal09,sal12,den16} or the cosmic evolution of the GRB rate \citep[e.g.,][]{but10,qin10,wp10}. Furthermore, \citet{sal12}, for example, suggested that a broken power law luminosity evolution with redshift is required to fit the observed redshift distribution.  Other researchers, however, proposed that it is not necessary to invoke luminosity evolution with redshift to explain the observed GRB rate at high z, by carefully taking selection effects into account \citep[e.g.,][]{wan13,den16}. Nevertheless, it is clear that any models that will resolve the properties of the GRBs, for self-consistency, must be able to fit the observed redshift and jet opening angle distributions at the same time for any instrument (pre-\Swift, \Swift, or \Fermi). To date, no other researchers have considered this method of analysis since the \citet{ld07} paper because there has always been some problem with estimating the achromatic jet break in the \Swift~data.

Fortunately, the \Swift~X-ray telescope (XRT) has detected nearly 700 X-ray afterglows through 2013~\citep[][]{bur05}, and these data were tested for jet break predictions \citep[][]{lia08,rac09}. \citet{van10} have suggested that jet breaks do exist in the \Swift~data and that they occur at a much later time (from minutes to  months). Furthermore, the GRB ejecta are not pointed directly at the observers, but at an off-axis angle $\theta_{obs}$, where $\theta_{obs} < \theta_{jet}$. \citet{zha15} further suggested that much later time observations are required to check whether jet breaks occur at or below the \Swift/XRT sensitivity limit of a few times $10^{-14}$ erg cm$^{-2}$ s$^{-1}$, and that these late-time and highly sensitive observations can only be carried out by {\it Chandra}, which has a limiting flux roughly an order of magnitude lower than the XRT for exposure times of order $60$ ks. Combining those {\it Chandra} data with well-sampled \Swift/XRT light curve observations and fitting the resultant light curves to numerical simulations, \citet{zha15} were able to estimate the jet opening angles for $27$ GRBs from the \Swift~sample (hereafter \Swift-Zhang). Moreover, \citet{rya15}, using an approach similar to that of \citet{zha15}, have estimated the jet opening angles for a complete sample size of more than 100 GRBs from the \Swift~data (hereafter \Swift-Ryan 2012). It is therefore interesting to revisit the work done by \citet{ld07} and reevaluate their model with more complete \Swift~redshift and jet opening angle distributions. However, it is important to mention that \citet{lia08} and \citet{wan15} have systematically investigated the jet-like breaks in the X-ray and optical afterglow light curves, and they showed that the optical ``jet break time''  agrees with those of X-rays in only a small fraction of bursts. Hence, in this work, we will predict the jet opening angle distributions for \Swift~and see how they fit the observed jet opening angle distributions from the \citet{zha15} and \citet{rya15} samples.

In this paper, we revisit the \citet{ld07} work by refitting the redshift and the jet opening angle distributions measured from both pre-\Swift~and \Swift~satellites with complete sample sizes. We further explore how the broken power law GRB spectrum affects the overall fitting of the redshift and the jet opening angle distributions by moving away from a flat GRB spectrum. The average beaming-corrected $\gamma$-ray energy released for pre-\Swift~and \Swift~is about $10^{51}$ erg \citep[e.g.,][]{blo03,fb05,cen10,zha15}, but others have also suggested that the average value for \Swift~could be an order of magnitude smaller than for pre-\Swift~\citep[e.g.,][]{kb08,gol16}, which could be due to \Swift's detection of lower-energy and higher-redshift events. However, in this work, we continue to assume a constant beaming-corrected energy release for all LGRBs to explore if such models could work with a modified jet opening angle distribution. We also examine the question of limited sample sizes. The paper is organized as follows. In Section~2 we briefly discuss the~\citet{ld07} cosmological GRB model and the differences between \Swift~and pre-\Swift~redshift and jet opening angle distributions. In Section~3 we discuss the results of our fit and conclude with a discussion of the derived GRB source rate density and the nature of the low- and high-redshift GRBs in Section~4. We adopt a $\Lambda$CDM cosmology in this paper, where the Hubble parameter is $H_0 = 72$ km s$^{-1}$ Mpc$^{-1}$, $\Omega_{m} = 0.27$, and $\Omega_{\Lambda} = 0.73$ \citep[e.g.][]{spe03}.

\section{Overview of the Cosmological GRB Model and the \Swift and Pre-\Swift~Redshift and Jet Opening Angle Distributions}
In this section we give an overview of the \citet{ld07} model and the modification that we intend to make in this paper, and for completeness, we restate the major equations from that paper here. We also give an overview of the complete redshift and observed estimated jet opening angle distributions from the \Swift~instruments.

\subsection{GRB Model}
\citet{ld07} approximated the spectral and temporal profiles of a GRB occurring at redshift $z$ by an emission spectrum that is constant for observing angles $\theta = \arccos\mu \leq \theta_{\rm j} = \arccos \mu_{\rm j}$ to the jet axis during the period $\Delta t_*$ with the $\nu F_\nu$ spectrum written as
\begin{equation}
\nu F_\nu \equiv \fluxeps(t) \cong f_{\epsilon_{pk}} S(x)
H(\mu; \mu_{\rm j},1) \ H(t;0, \Delta t) \ . \\
\label{eq1}
\end{equation}
Here, the spectral energy density (SED) function $S(x) = 1$ at $x = \e/\e_{pk} = \e_*/\e_{pk*}$, and the duration of the GRB in the observer's frame is $\Delta t = (1+z) \Delta t_*$ (stars refer to the stationary frame, and terms without stars refer to observer quantities). In this work, similar to \citet{ld07}, we take $\Delta t_* = 10$ s because this is the mean value of the GRB duration measured with BATSE assuming that BATSE GRBs are typically at $z \approx 1$. At observer time $t$, $\epsilon = h\nu/m_e c^2$ is the dimensionless energy of a photon in units of the electron rest-mass energy, and $\epsilon_{pk}$ is the photon energy at which the energy flux $f_\epsilon $ takes its maximum value $f_{\epsilon_{pk}}$.  The quantity $H(\mu; \mu_{\rm j}, 1)$ is the
Heaviside function such that $H(\mu; \mu_{\rm j}, 1) = 1$ when $\mu_{\rm j} \leq \mu \leq 1$ (or when the angle $\theta $ of the
observer with respect to the jet axis is within the opening angle of the jet), and $H(\mu; \mu_{\rm j}, 1) = 0$ otherwise.  One possible approximation to the GRB spectrum is a broken power law, so that
\begin{equation}
S(x) = \left[ \left(\frac{\epsilon}{\epsilon_{pk}}\right)^a
H(\epsilon_{pk} - \epsilon) +
\left(\frac{\epsilon}{\epsilon_{pk}}\right)^b H(\epsilon -
\epsilon_{pk}) \right] \ , \\
\label{eq2}
\end{equation}
where the Heaviside function $H(u)$ of a single index is defined such that $H(u) = 1$ when $x \geq 0$ and $H(u) = 0 $ otherwise. The $\nu F_\nu$ spectral indices are denoted by $a(\geq0)$ and $b(\leq0)$. \citet{ld07} assumed a flat spectrum, so $a=b=0$ in their work.

The bolometric fluence of the model GRB for observers with $\theta \leq \theta_{\rm j}$ is given by
\begin{equation}
F = \int^\infty_{-\infty} dt \; \int^\infty_0 d\e\; \frac{\fluxeps(t)
}{\epsilon}  = \lambda_b\; f_{\epsilon_{pk}}\;\Delta t  \;,
\label{eq3}
\end{equation}
where $\lambda_b$ is a bolometric correction to the peak measured $\nu F_\nu$ flux. If the SED is described by Equation~(\ref{eq2}), then $\lambda_b = (a^{-1} - b^{-1})$ and is independent of $\e_{pk}$. Note, if $a=b=0$, then $\lambda_b \longrightarrow ln(x_{\rm max}/x_{\rm min})$, and since the model spectrum is not likely to extend beyond more than two orders of magnitude, $\lambda_b \lesssim 5$; hence, \citet{ld07} take $\lambda_b = 5$ in their calculation. The spectral power-law indices $a$ and $b$ are related to the Band function, which relates the low-energy index $\alpha \approx -1$ and the high-energy index $\beta \approx -2.5$ as the generally accepted values \citep[e.g.][]{ban93,pre00}. In this work, we use $\alpha = -1$ and $\beta = -2.5$, which imply the spectral power-law indices $a=1$ and $b = -0.5$, respectively; thus, we have $\lambda_{b} = 3$ as the bolometric correction constant. 

The beaming-corrected $\gamma$-ray energy release $\Estarg$ for a two-sided jet is given by
\begin{equation}
\Estarg = 4 \pi d^2_L(z) (1-\mu_{\rm j}) \; \frac{F}{1+z}  \ , \\
\label{eq4}
\end{equation}
where the luminosity distance
\begin{equation}
d_L(z) = \frac{c}{H_0}(1+z) \ \int^z_0
\frac{dz^\prime}{\sqrt{\Omega_m (1+z^\prime)^3 + \Omega_\Lambda}}
\label{eq5}
\end{equation}
for a $\Lambda$CDM universe, where $c$, $H_0$, $z$, $\Omega_m$, and $\Omega_\Lambda$ are the light speed, the current Hubble constant, the distance redshift, and the dark matter and dark energy densities, respectively.  Substituting Equation~(\ref{eq3}) for $F$ into Equation~(\ref{eq4}) gives the peak flux
\begin{equation}
f_{\epsilon_{pk}} = \frac{\Estarg}{4 \pi d^2_L(z) (1 - \mu_{\rm j})
\Delta t_* \ \lambda_b} \ . \\
\label{eq6}
\end{equation}
Finally, by substituting Equation~(\ref{eq6}) into Equation~(\ref{eq1}), the energy flux becomes
\begin{eqnarray}
\fluxeps(t)  \; = \; \frac{\Estarg \;H(\mu; \mu_{\rm j},1) \ H(t;0,
\Delta t) S(x)}{4 \pi d^2_L(z) (1 - \mu_{\rm j}) \Delta t_* \lambda_b}
\;. \label{eq7}
\end{eqnarray}
Clearly, Equations~(\ref{eq4}), (\ref{eq6}), (\ref{eq7}), and the bursting rate equations, which will be discussed below, are affected by the new bolometric correction constant.

\subsection{Bursting Rate of GRB Sources}

To calculate the redshift, size, and the jet opening angle distributions, we use Equations (16), (18), and (20) from \citet{ld07}, which describe the directional GRB rate per unit redshift ($d\dot{N}(> \fluxthres)/d\Omega dz$) with energy flux $>\fluxthres$, the size distribution of GRBs ($d\dot{N}(> \fluxthres)/d\Omega$) in terms of their $\nu F_\nu$ flux $f_\e$, and the observed directional event reduction rate (due to the finite jet opening angle) for bursting sources with $\nu F_\nu$ spectral flux greater than $\fluxthres$ at observed photon energy $\e$ or simply the jet opening angle distribution ($d\dot{N}(> \fluxthres)/d\Omega d\mu_{\rm j}$), respectively. These equations are given by
\begin{eqnarray}
\frac{d\dot{N}(> \fluxthres)}{d\Omega \ dz }  =  \frac{c g_0}{H_0
(2+s)} \frac{d^2_L(z) \ \dot{n}_{co}(z) }{(1+z)^3 \ \sqrt{\Omega_m
(1+z)^3 + \Omega_\Lambda}} \;
 \{ [1-\max(\hat\mu_{\rm j},\mu_{\rm j,min})]^{2+s} - (1-\mu_{\rm j,max})^{2+s}\}\;,
\label{eq8}
\end{eqnarray}

$$\frac{d\dot{N}(> \fluxthres)}{d\Omega  }  =
\frac{c g_0}{H_0
(2+s)} \int_0^{\zmax} dz\;\frac{d^2_L(z) \ \dot{n}_{co}(z)
}{(1+z)^3 \ \sqrt{\Omega_m (1+z)^3 + \Omega_\Lambda}} \;$$
\begin{eqnarray}
 \times \{ [1-\max(\hat\mu_{\rm j},\mu_{\rm j,min})]^{2+s} - (1-\mu_{\rm j,max})^{2+s}\}\;,
\label{eq9}
\end{eqnarray}
and
\begin{equation}
\frac{d\dot{N}(> \fluxthres)}{d\Omega d\mu_{\rm j}} = \frac{c}{H_0}
g(\mu_{\rm j})(1-\mu_{\rm j}) \int_0^{\zmax(\mu_{\rm j})}dz\; \frac{d^2_L(z) \
\dot{n}_{co}(z)}{(1+z)^3 \ \sqrt{\Omega_m (1+z)^3 +
\Omega_\Lambda}} \;, 
\label{eq10}
\end{equation}

\noindent where $f_\epsilon$ is given by Equation~(\ref{eq7}), $g(\mu_{\rm j})$ is the jet opening angle distribution, $\dot{n}_{co}(z)$ is the comoving GRB rate density, and $\fluxthres$ is the instrument's detector sensitivity. We take the energy flux $\fluxthres \sim 10^{-8} \ \rm erg \ cm^{-2} \ s^{-1}$ and $\sim 10^{-7} \ \rm erg \ cm^{-2} \ s^{-1}$ as the effective \Swift~and pre-\Swift~detective flux thresholds, respectively~\citep[see][and references therein]{ld07}. Since the form for the jet opening angle $g(\mu_{\rm j})$ is unknown, we also consider the function
\begin{equation}
g(\mu_{\rm j}) = g_0 \ (1-\mu_{\rm j})^s \ H(\mu_{\rm j};\mu_{\rm j,min},\mu_{\rm j,max}) \;,
\label{eq11}
\end{equation}
where $s$ is the jet opening angle power-law index; for a two-sided jet, $\mu_{\rm j,min} \geq 0$, and 
\begin{equation}
g_0 = \frac{1+s}{(1-\mu_{\rm j,min})^{1+s} - (1-\mu_{\rm j,max})^{1+s}} \;
\label{eq12}
\end{equation}
is the distribution normalization~\citep{ld07}. This functional form $g(\mu_{\rm j})$ describes GRBs with small opening angles, that is, with $\theta_{\rm j} \ll 1$, that will radiate their available energy into a small cone, so that such GRBs are potentially detectable from larger distances with their rate reduced by the factor $(1-\mu_{\rm j})$. By contrast, GRB jets with large opening angles are more frequent, but only detectable from comparatively small distances~\citep[e.g.,][]{gpw05,ld07}. From \citet{ld07} Equations~(17) and (19), we have 
\begin{equation}
\hat\mu_{\rm j} \; \equiv \; 1 - \frac{\Estarg}{4 \pi d^2_L(z) \ \Delta
t_* \ \fluxthres \lambda_b}\;
\label{eq13}
\end{equation}
and the value of the maximum redshift $z_{\rm max}$, the integral limit in Equations~(\ref{eq9}) and (\ref{eq10}), is obtained by satisfying the condition
\begin{eqnarray}
d^2_L(\zmax) & \leq & \frac{\Estarg}{4 \pi (1 - \mu_{\rm j,min}) \ 
\Delta t_* \ \fluxthres \lambda_b}  \ . 
\label{eq14}
\end{eqnarray}
\begfig[t] \hskip-0.25in \epsscale{1.15} \plottwo{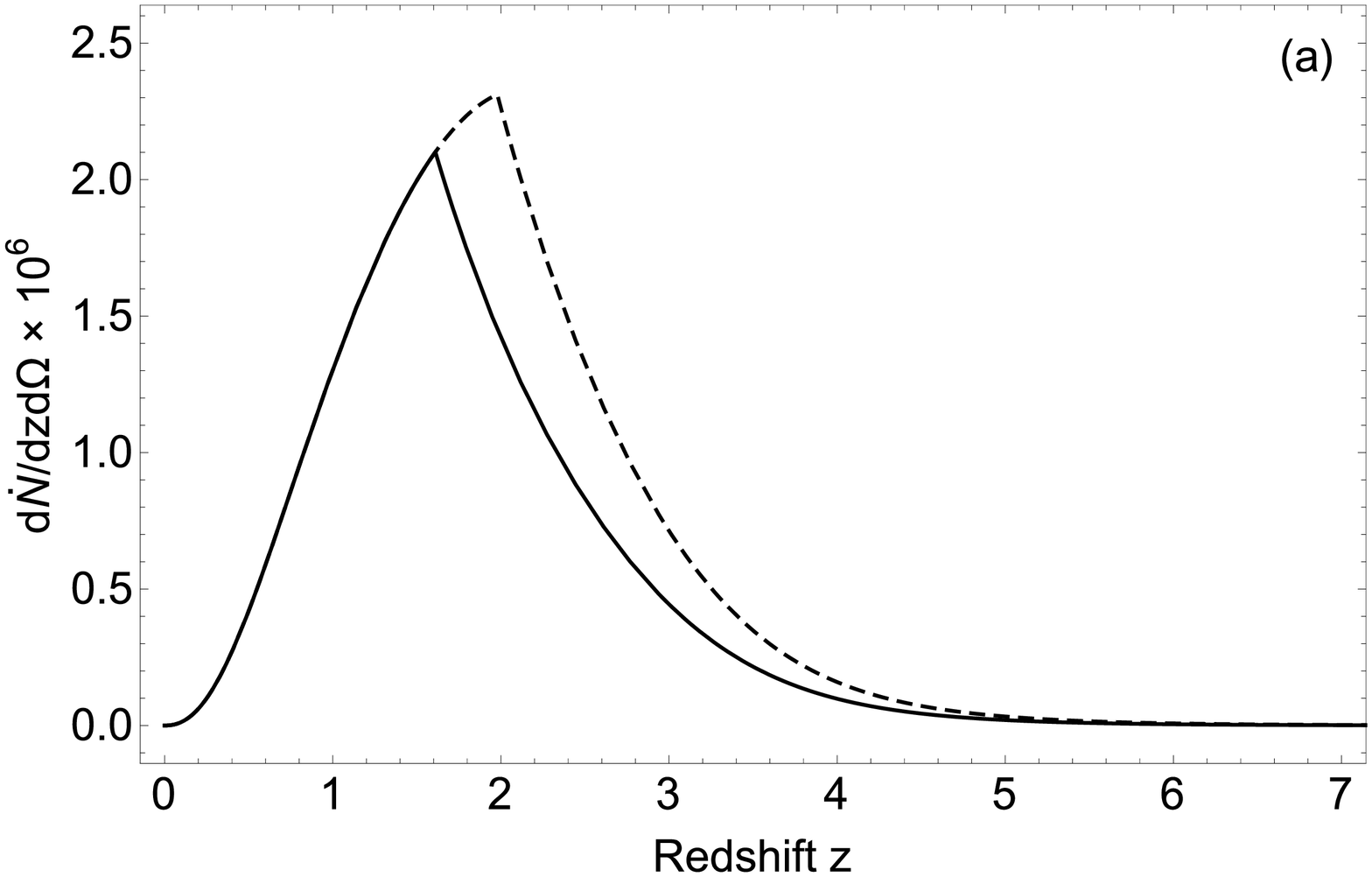}{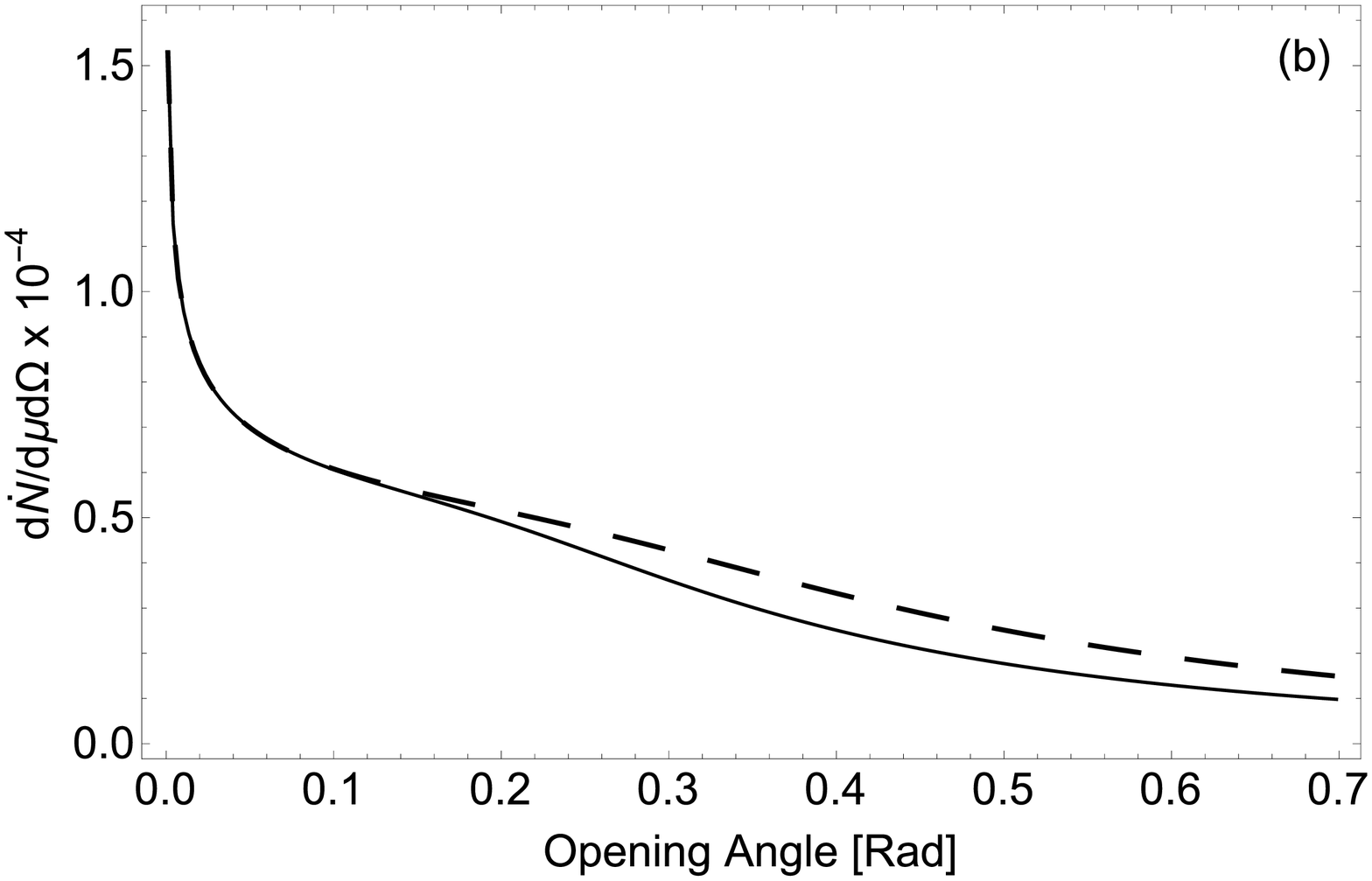}
\caption{\footnotesize (a) Directional event rate per unit redshift (one event per $(10^{28} \ \rm cm)^3$ per day per sr per $z$) and (b) directional event rate per unit jet opening angle (one event per $(10^{28} \ \rm cm)^3$ per day per sr per $\mu_{\rm j}$) using $\theta_{\rm j,min} = 0.05 \ \rm rad$ and $\theta_{\rm j,max} = 0.7 \ \rm rad $,  $s = -1.1$, $\Estarg = 2 \times 10^{52} \ \rm erg$, $\fluxthres = 10^{-7} \ \rm erg \ cm^{-2} \ s^{-1}$, and SFR3. The solid and dashed curves represent the bolometric corrections $\lambda_b = 5$ and $3$, respectively.}
\label{fig2} 
\finfig
For demonstration purposes, the plots of the directional event rate per unit redshift per steradian (see Equation~(\ref{eq8})) and directional event rate per unit jet opening angle per steradian (see Equation~(\ref{eq10})) are depicted in Figure~\ref{fig2} for bolometric correction constants $\lambda_{b} = 5$ and $3$. As we move away from the flat spectrum, the bolometric correction $\lambda_b$ gets smaller, and this reduces the beaming-corrected $\gamma$-ray energy release $\Estarg$ (see Equation (\ref{eq4})) but enhances the energy flux (see  Equation (\ref{eq7})).  As a  result, the directional event rate per unit redshift per steradian and directional event rate per unit jet opening angle per steradian are affected at low redshift and large jet opening angle as depicted in Figure~\ref{fig2}. These results suggest that we will expect to see overall increases in the total number of GRBs when we move away from the flat spectrum.  

Finally, in this work we also assume the comoving GRB rate density to be 
\begin{equation}
\dot{n}_{co}(z) = \dot{n}_{co} \Sigma_{_{\rm SFR}}(z) \;,
\label{eq15}
\end{equation}
where $\dot{n}_{co}$ is the comoving rate density normalization constant, and $\Sigma_{_{\rm SFR}}$ is the GRB formation rate functional form. It has been suggested~\citep[e.g.,][]{tot97,nat05} that the GRB formation history is expected to follow the cosmic SFR derived from the blue and UV luminosity density of distant galaxies, but \citet{ld07}, for example, have shown that in order to obtain the best fit to the \Swift~redshift and the pre-\Swift~redshift and jet opening angle samples, they must employ a GRB formation history that displays a monotonic increase in the SFR at high redshifts~\citep[SFR5 and SFR6 models; see Fig.~\ref{fig1}(a) here or Fig.~4 in][]{ld07}. \citet{ld07} utilized the GRB formation rate based on the observed SFR history of~\citet{hb06}, which is given as
\begin{equation}
\Sigma_{_{\rm SFR}}(z) = \frac{1 + (a_2 z/a_1)}{1+ (z/a_3)^{a_4}} \;, 
\label{eq16}
\end{equation}
where $a_1 = 0.015$, $a_2 = 0.10$, $a_3 = 3.4$, and $a_4 = 5.5$ are their best-fits parameters, and $a_1=0.015$, $a_2=0.12$, $a_3=3.0$, $a_4=1.3$, and $a_1=0.011$, $a_2=0.12$, $a_3=3.0$, $a_4=0.5$ are the~\citet{ld07} SFR5 and SFR6 models, respectively, which give a good fit to the \Swift~and pre-\Swift~redshift distributions~\citep{ld07}.  In this paper we also examine the GRB formation rate of the forms 
\begin{equation}
    \Sigma_{_{\rm SFR}}(z)= \frac{(1+a_1)}{(1+z)^{-a_2} + a_1 (1+z)^{a_3}}
    \label{eq17}
\end{equation}
and
\begin{equation}
    \Sigma_{_{\rm SFR}}(z)=\left\{
                \begin{array}{ll}
                  a_0 (1+z)^\alpha \hskip+0.25in , 0 \leq z \leq z_1 \\
                  b_0 (1+z)^\beta \hskip+0.25in , z_1 < z \leq z_2  \;, \\
                  c_0 (1+z)^\gamma \hskip+0.25in , z > z_2 
                \end{array}
              \right.  
\label{eq18}
\end{equation}
where $a_1, a_2, a_3, a_0, b_0, c_0, \alpha, \beta$, and $\gamma$ are constants; and we set $a_0 = 1$ and $b_0 = a_0 (1+z)^\alpha / (1+z)^\beta$ and $c_0 = b_0 (1+z)^\beta / (1+z)^\gamma$ to ensure continuity at $z_1$ and $z_2$, respectively. The constant values of $a_1, a_2$, and $a_3$ in Equation (\ref{eq17}) and $\alpha, \beta$, and $\gamma$ in Equation (\ref{eq18}) are constrained by fitting the pre-\Swift~and \Swift~redshift and jet opening angle distributions.

Since it is unclear whether the excess of GRB rate at high redshift is due to luminosity evolution or some other means, or that GRBs rates do simply follow the SFR, in this paper we continue to search for the GRB formation functional form that best fits the \Swift~and pre-\Swift~redshift and jet opening angle distributions at the same time. We expect the GRB rate functional form will be different from the result obtained by~\citet{ld07} because the redshift, size, and opening angle distributions from Equations~(\ref{eq8})-(\ref{eq10}), respectively, are affected by the new bolometric correction constant that we discussed in Section~2.1, and also because we now have more complete \Swift~redshift and jet opening angle samples, as discussed below. 

\subsection{\Swift~and Pre-\Swift~Redshift and Jet Opening Angle Distributions}

With the launch of the \Swift~satellite~\citep{geh04}, rapid follow-up studies of GRBs triggered by the Burst Alert Telescope (BAT) on \Swift~became possible. A fainter and more distant population of GRBs than found with the pre-\Swift~satellites CGRO-BATSE, BeppoSAX, INTEGRAL, and HETE-2 is detected~\citep{ber05}. Before 2008, the mean redshift of pre-\Swift~GRBs that also have measured beaming breaks~\citep{fb05} is $\langle z\rangle \sim 1.5$, while GRBs discovered by \Swift~have $\langle z\rangle \sim 2.7$~\citep{jak06}. However, insufficient observed estimated jet opening angles were obtained to produce a statistically reliable sample for \Swift~because of problems with estimating the achromatic jet breaks in the data. From~\citet{fb05}, the pre-\Swift~jet opening angle distribution extends from $\theta_{\rm j} = 0.05$ to $\theta_{\rm j} = 0.6$ rad. However, other workers have reported that the observed jet opening angles are as large as $\theta_{\rm j} = 0.7 \ \rm rad $ \citep[e.g.,][]{gpw05}. \citet{ld07} showed that the best fit for the \Swift~redshift and pre-\Swift~redshift and opening angle distributions, assuming a flat spectrum, required $\theta_{\rm j} = 0.05$ to $\theta_{\rm j} = 0.7$ rad. 

For more than a decade, the accumulation of data for the redshift and opening angle from the pre-\Swift~instruments has not increased in sample size because most of the pre-\Swift~instruments have been decommissioned, except INTEGRAL. However, for the \Swift~instrument, the redshift sample size with known redshift has increased to more than $53$ GRBs~\citep[e.g.,][]{jak12,rya15}. More importantly, the jet opening angles from the \Swift~sample are still lacking because of the missing achromatic jet break time~\citep[e.g.,][]{kz15,zha15}. Fortunately, \citet{zha15} were able to estimate the jet opening angles for $27$ GRBs from the \Swift~sample by combining late-time {\it Chandra} data with well-sampled \Swift/XRT light curve observations and by fitting the resultant light curves to the numerical simulation assuming an off-axis angle. Figures~\ref{fig3}a-\ref{fig3}c depict the redshift and jet opening angle distributions from the \Swift-Zhang~\citep{zha15} and pre-\Swift~\citep{fb05} samples. 

Other researchers, for example, \citet{wan13}, \citet{den16}, and \citet{jap16}, have suggested that the study of GRB rate is not sufficient if the GRB sample size is incomplete. Fortunately, \citet{rya15}, using an approach similar to that of \citet{zha15}, have estimated the jet opening angles of more than $100$ GRBs between 2005 and 2012 (\Swift-Ryan 2012). However, only 15 sources with ``well-fit'' light curves and observed jet opening angles and 32 sources (hereafter \Swift-Ryan-b) with ``well-fit'' jet opening angles are constrained.  Interestingly, the \Swift-Ryan-b sample gives a redshift distribution similar to that of the  \citet{zha15} sample, but their mean jet opening angle (e.g., $\langle \theta_{\rm j} \rangle \sim 0.1$ rad) distribution is much smaller that that of the \citet{zha15} sample (e.g., much greater than $0.1$ rad), as shown in Figure~\ref{fig3}(c). This result could be related to the fact that the \Swift-Zhang sample is a subset of the \Swift-Ryan 2012 sample, where most of the Chandra-observed GRBs are at the bright end of the whole \Swift~ GRB sample \citep[][]{zha15}. Figure~\ref{fig3}(d) is the cumulative jet opening angle distribution for the complete \Swift-Ryan 2012 sample. The shaded regions in Figures~\ref{fig3}(c) and \ref{fig3}(d) are the statistical error bars for the \Swift-Zhang and \Swift-Ryan 2012 samples, respectively. The \Swift-Ryan-b sample has much smaller error bars, but we do not plot them here to make it easier to read the curves in Figure~\ref{fig3}(c). In Figures~\ref{fig3}(a)-\ref{fig3}(d), it is also interesting to note that the \Swift-Zhang, \Swift-Ryan-b, and \Swift-Ryan-2012 samples all have the same median redshift value, $\langle z \rangle \sim 1.7$, but their associated median jet opening angles are anywhere in the range $\langle \theta_{\rm j} \rangle \sim [0.1 - 0.35]$ rad; these indicate that there is more work to be done in constraining the jet opening angles from the \Swift~data. In this study, we also explore a possible GRB burst rate functional form by fitting the current complete estimated \Swift-Ryan 2012~and pre-\Swift~redshift and jet angle distributions from \citet{rya15} and \citet{fb05}, respectively. It is important to realize that an acceptable GRB burst rate functional form is one that gives acceptable fits to both the \Swift~and pre-\Swift~redshift and jet opening angle distributions at the same time. 
\begfig[t] \hskip-0.25in \epsscale{1.15} \plottwo{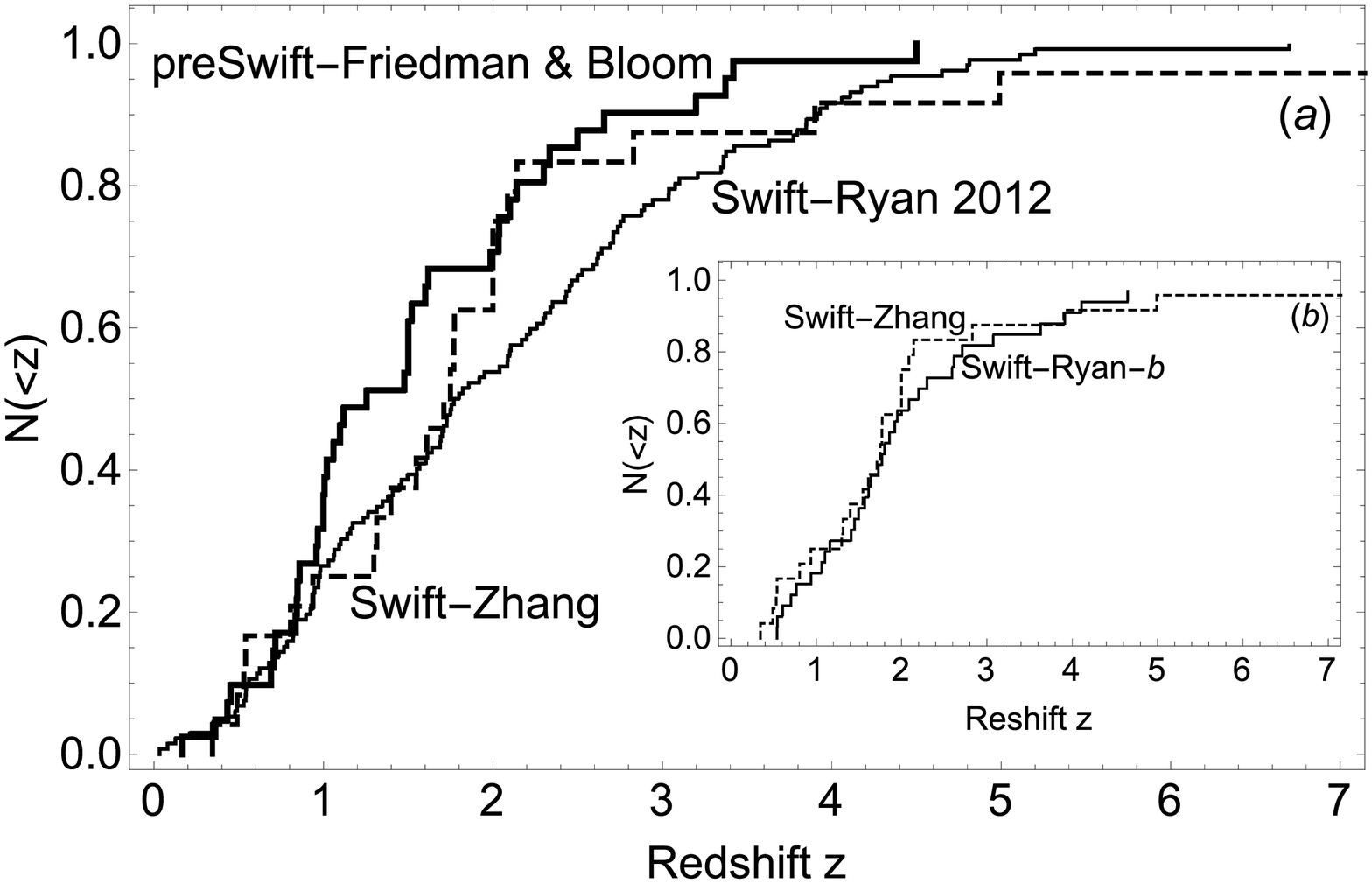}{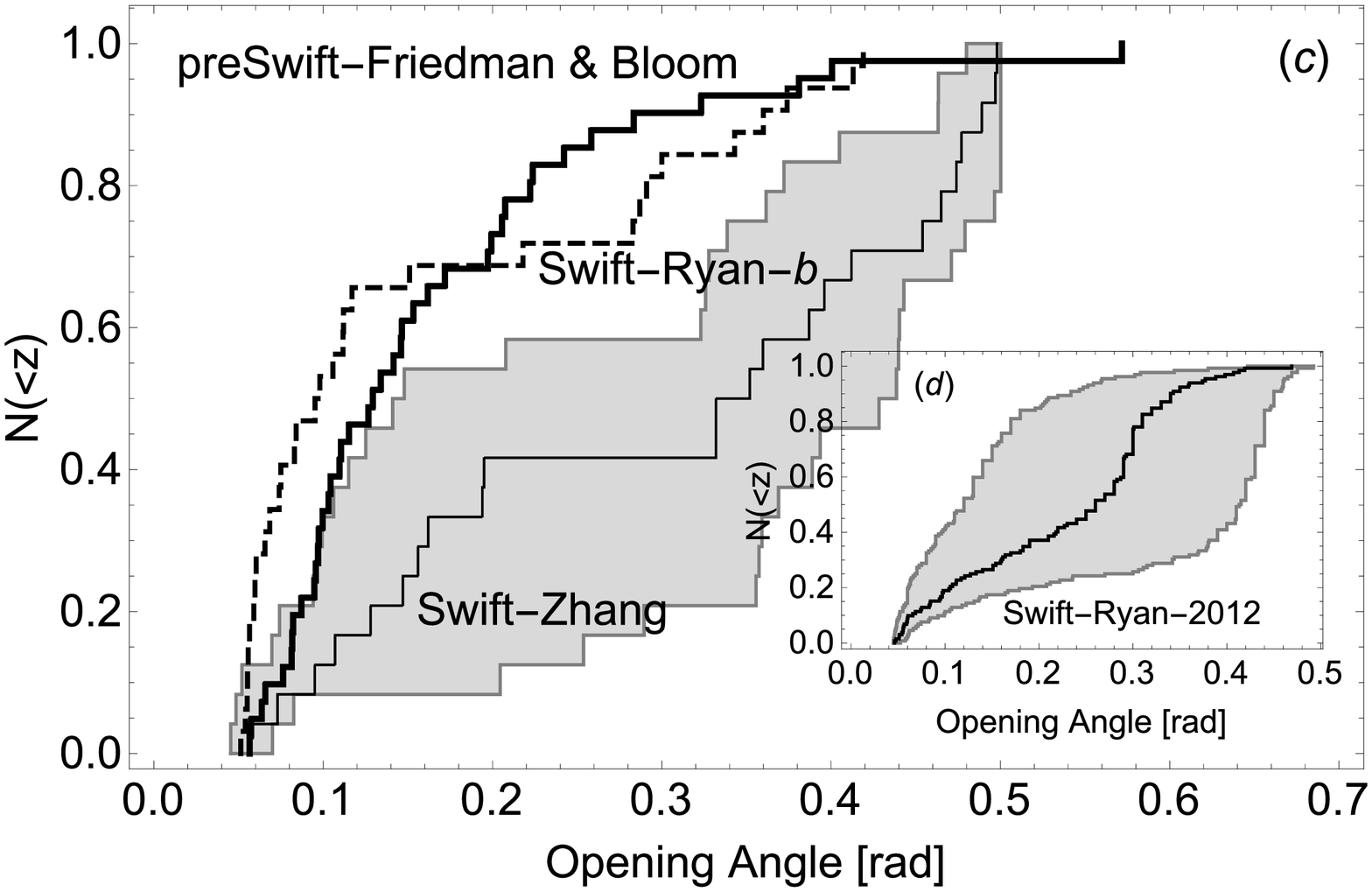}
\caption{\footnotesize (a) \Swift-Zhang and \Swift-Ryan-b are incomplete GRB samples. \Swift-Ryan-2012 and pre-\Swift~are complete GRB samples. Plotted are the cumulative redshift distributions of 41 GRBs in the pre-\Swift~sample~\citep[thick solid line;][]{fb05}, 27 GRBs in the \Swift-Zhang sample~\citep[dashed line;][]{zha15}, and 133 GRBs in the \Swift-Ryan-2012 sample~\citep[thin solid line;][]{rya15}. (b) Cumulative redshift distribution of 33 GRBs in the \Swift-Ryan-b sample~\citep[``well-fit'' observed jet opening angles; thick solid line;][]{rya15}  and 27 GRBs in the \Swift-Zhang sample. The median redshifts of the pre-\Swift~and \Swift~bursts are $\langle z \rangle \sim 1.2$ and $\langle z \rangle \sim 1.7$, respectively. (c) The cumulative opening angle distributions of 41 GRBs in the pre-\Swift~sample~\citep[thick solid line;][]{fb05}, 27 GRBs in the \Swift-Zhang sample~\citep[thin solid line;][]{zha15}, and 33 GRBs in the \Swift-Ryan-b sample~\citep[dashed line;][]{rya15}. The median opening angles in the pre-\Swift, \Swift-Ryan-b, and \Swift-Zhang~samples are $\langle \theta_{jet} \rangle \sim 0.13, \, \sim 0.1$, and $> 0.1$ rad, respectively. The shaded region is the estimated error bars from the \Swift-Zhang sample. The \Swift-Ryan-b jet opening angle sample also has the estimated error bars but are much tighter in comparison to the \Swift-Zhang estimates (and they are not shown here for clarity). (d) The cumulative opening angle distribution of 133 GRBs in the \Swift-Ryan 2012  sample~\citep[solid line;][]{rya15}. The shaded region is the estimated error bars for the complete \Swift-Ryan 2012 sample.}
\label{fig3} 
\finfig
\begfig[t] \hskip-0.25in \epsscale{1.15} \plottwo{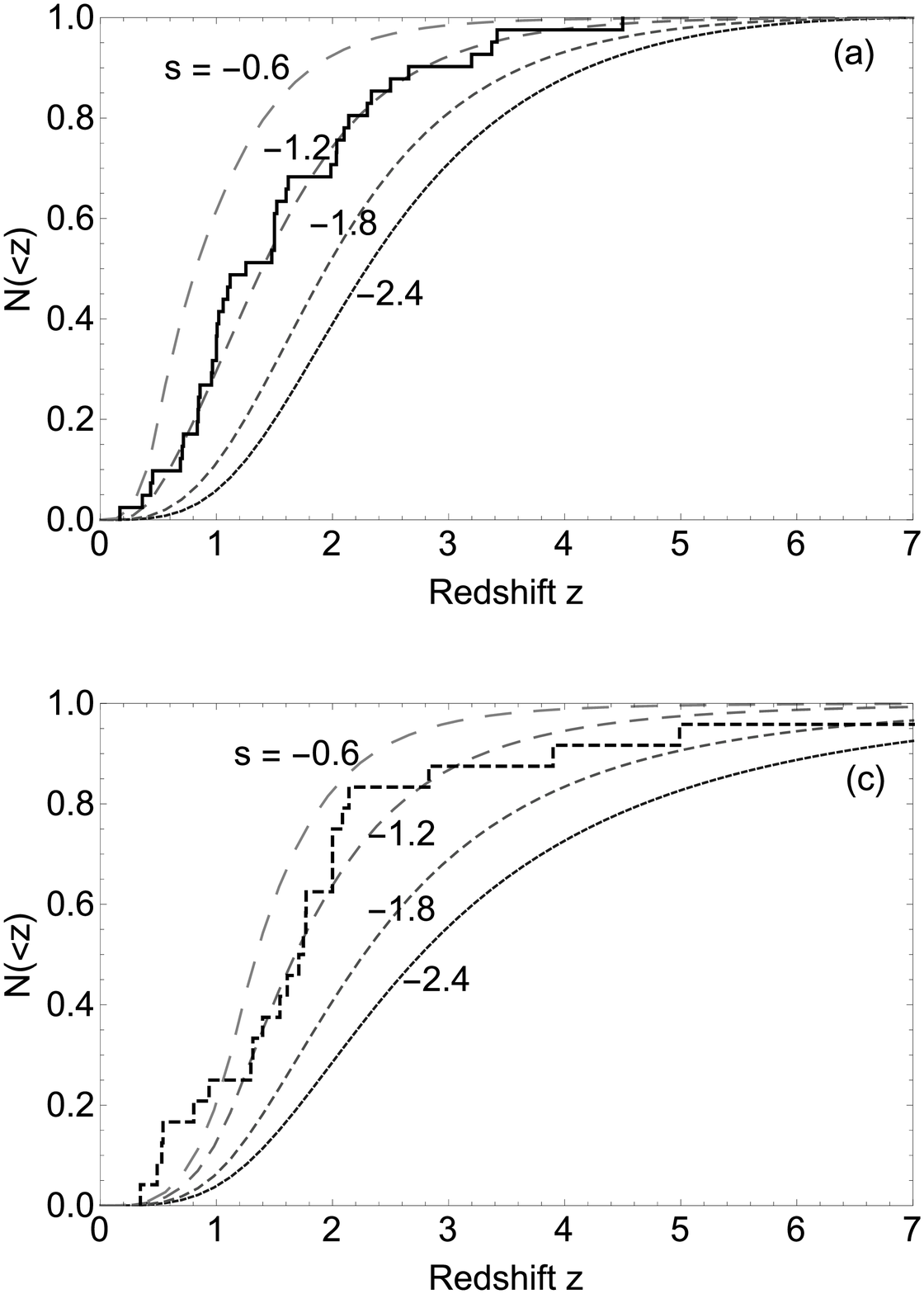}{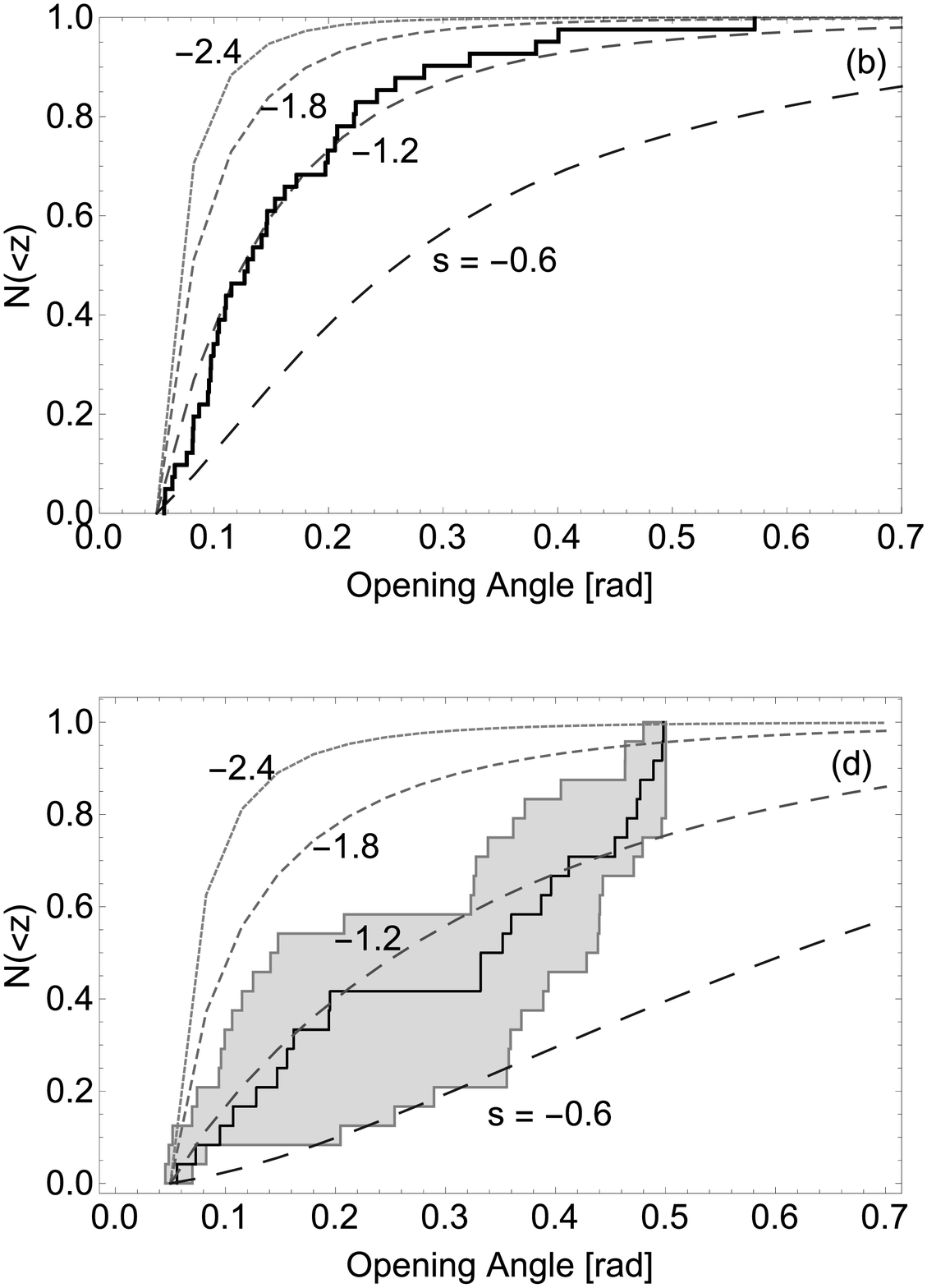}
\caption{\footnotesize (a) The cumulative redshift and (b) jet opening angle distributions of the fitted model (thin dashed line) and the pre-\Swift~sample~\citep[thick line;][]{fb05}. (c) The cumulative redshift and (d) jet opening angle distributions of the fitted model and the \Swift-Zhang~sample~\citep[thick line;][]{zha15}. These fitted results assume a flat spectrum with SFR8 functional form, the range of jet opening angles $\theta_{\rm j,min} = 0.05$ rad to $\theta_{\rm j,max} = 1.57$ rad, the jet opening angle power-law index $s = -0.6, -1.2, -1.8$, and $-2.4$, and the $\gamma$-ray energy released $\Estarg = 4 \times 10^{51}$ erg.}
\label{fig4} 
\finfig

\section{Results And Discussion}
Similar to the work done by~\citet{ld07}, this model has seven adjustable parameters: the $\nu F_\nu$ spectral power-law indices $a$ and $b$, the power-law index $s$ of the jet opening angle distribution, the range of the jet opening angles $\theta_{\rm j,min}$ and $\theta_{\rm j,max}$, the average absolute emitted $\gamma$-ray energy $\Estarg$, and the detector threshold $\fluxthres$. As already mentioned, we consider a broken power law $\nu F_\nu$ GRB SED with $a =1$ and  $b = -0.5$ as generally accepted values, which gives the bolometric correction factor $\lambda_b = 3$. The flux thresholds $\fluxthres$ are set equal to $10^{-8}$ erg cm$^{-2}$ s$^{-1}$ and $10^{-7}$ erg cm$^{-2}$ s$^{-1}$ for \Swift~and pre-\Swift, respectively.  The remaining parameters $\theta_{\rm j,min}$, $\theta_{\rm j,max}$, $s$, $\Estarg$, and the GRB rate functional form are constrained by obtaining the best fits to the redshift and jet opening angle distributions for both \Swift~and pre-\Swift~instruments.

\subsection{Possible GRB Formation Rates}
In our first analysis, we fit the new data, \Swift-Zhang~\citep{zha15}, \Swift-Ryan-b~\citep{rya15}, and pre-\Swift~\citep{fb05} using the flat GRB spectrum. The results of the fits are in Figures~\ref{fig4}(a)-\ref{fig4}(d) with the associated GRB density burst rate SFR8 (using Equation~(\ref{eq17}) with $a_1 = 0.005, a_2 = 4.5,$ and $a_3 = 1$; see Figure~\ref{fig1}(b)). The best-fit parameters have the range of jet opening angles $\theta_{\rm j,min} = 0.05$ rad to $\theta_{\rm j,max} = 1.57$ rad, the jet opening angle power-law index $s = -1.2$, and the $\gamma$-ray energy released $\Estarg = 4 \times 10^{51}$ erg, which is a factor of 2 larger than the observed average $\gamma$-ray energy released \citep[see][and references therein]{ld07}.  We also examine different values of the jet opening angle power-law index $s$ (see Figures~\ref{fig3}(a)-\ref{fig3}(d)), but only $s=-1.2$ provides the best results. The one-sample Kolmogorov-Smirnov (KS-1) test indicates that the observed estimated \Swift-Zhang, \Swift-Ryan-b, and pre-\Swift~redshift and jet opening angle distributions and their associated fit probability statistics (p-statistics) are greater than $0.05$, so the null hypothesis is rejected, indicating that the samples and their associated fits belong to the same distribution. The results from this analysis suggest that the GRB density burst rate (SFR8) does decline at high redshift, a totally different conclusion than that reached by \citet{ld07}. However, the suggested GRB burst rate (SFR8) from this analysis is not the same as any of the observed SFRs (see Figure~\ref{fig1}(a)). Moreover, it is unlikely that the jet opening angles for any of the GRBs will be as large as $1.57$ rad, since the current range of the observed estimated jet opening angles is between $0.05-0.7$ rad \citep[e.g.][]{fb05,gpw05,rya15,zha15}. Furthermore, we notice that the calculated \Swift~jet opening angle distribution from our model gives a better fit to the \Swift-Zhang distribution than to the \Swift-Ryan-b distribution (see Figures~\ref{fig3}(c) and~\ref{fig4}(d)). 

Moving away from the flat spectrum, we fit the \Swift-Zhang, \Swift-Ryan-b, and pre-\Swift~samples using the broken power law spectrum with $a = 1$ and $b = -0.5$ as the low- and high-energy indices, respectively. After examining the fitted results from different parameters and SFR functional forms (see Equations~(\ref{eq16})-(\ref{eq18})) that could affect the outcome of the fits using our model, we realize that it is possible to fit the observed \Swift-Zhang, \Swift-Ryan-b, and pre-\Swift~redshift and jet opening angle distributions using the GRB rate functional form SFR9 (using Equation~(\ref{eq18}) with $\alpha = 4.1, \beta = 0.8, \gamma = -5.1, z_1 = 0.5,$ and $z_2 = 4.5$; see Figure~\ref{fig1}(b)), similar to that of~\citet{hb06} and as extended by~Li (2008; see SFR7 in Figures~\ref{fig1}(a) and \ref{fig1}(b)). The fitted results are shown in Figure~\ref{fig5} for the redshift and jet opening angle distributions for the pre-\Swift, \Swift-Zhang, and \Swift-Ryan-b samples. The one-sample KS-1 test indicates that the observed estimated \Swift-Zhang, \Swift-Ryan-b, and pre-\Swift~redshift and jet opening angle distributions and their associated fit p-statistics are greater than $0.05$, so the null hypothesis is rejected. The best-fit parameters have the range of jet opening angles between $\theta_{\rm j,min} = 0.05$ rad and $\theta_{\rm j,max} = 0.7$ rad, the jet opening angle power-law index $s = -1.2$, and the $\gamma$-ray energy released $\Estarg = 2 \times 10^{51}$ erg.  For the first time, our results show, self-consistently, that GRBs do follow the observed star formation history similar to SFR7 using the \Swift-Zhang~\citep{zha15}, \Swift-Ryan-b~\citep{rya15}, and pre-\Swift~\citep{fb05} samples. Our fitted result of the jet opening angle distribution for the \Swift~instrument is between the estimated~\Swift-Zhang and \Swift-Ryan-b~samples (see Figure~\ref{fig5}(d)), and it is also within the error bars of the \citet{zha15} estimated values. Furthermore, we notice that the fit to the jet opening angle distribution is more consistent with the observed estimated jet opening angle distribution from the \Swift-Zhang sample (see Figures \ref{fig4}(d) and \ref{fig5}(d)). Moreover, when we change the GRB functional form power-law index (see Equation~(\ref{eq18})) from $\gamma = -5.1$ to $\gamma = -2.0$ (more than three orders of magnitude) beyond $z > 4.5$ with other power-law indices ($\alpha = 4.1$ and $\beta = 0.8$) remaining the same, the fits show very little variation in the calculated \Swift~redshift and jet opening angle distributions and almost none for the calculated pre-\Swift~distributions; this is expected for pre-\Swift, since its detection threshold is insensitive at high redshift. These results suggest that the GRB rate does follow the SFR at high redshift. 
\begfig[t] \hskip-0.25in \epsscale{1.15} \plottwo{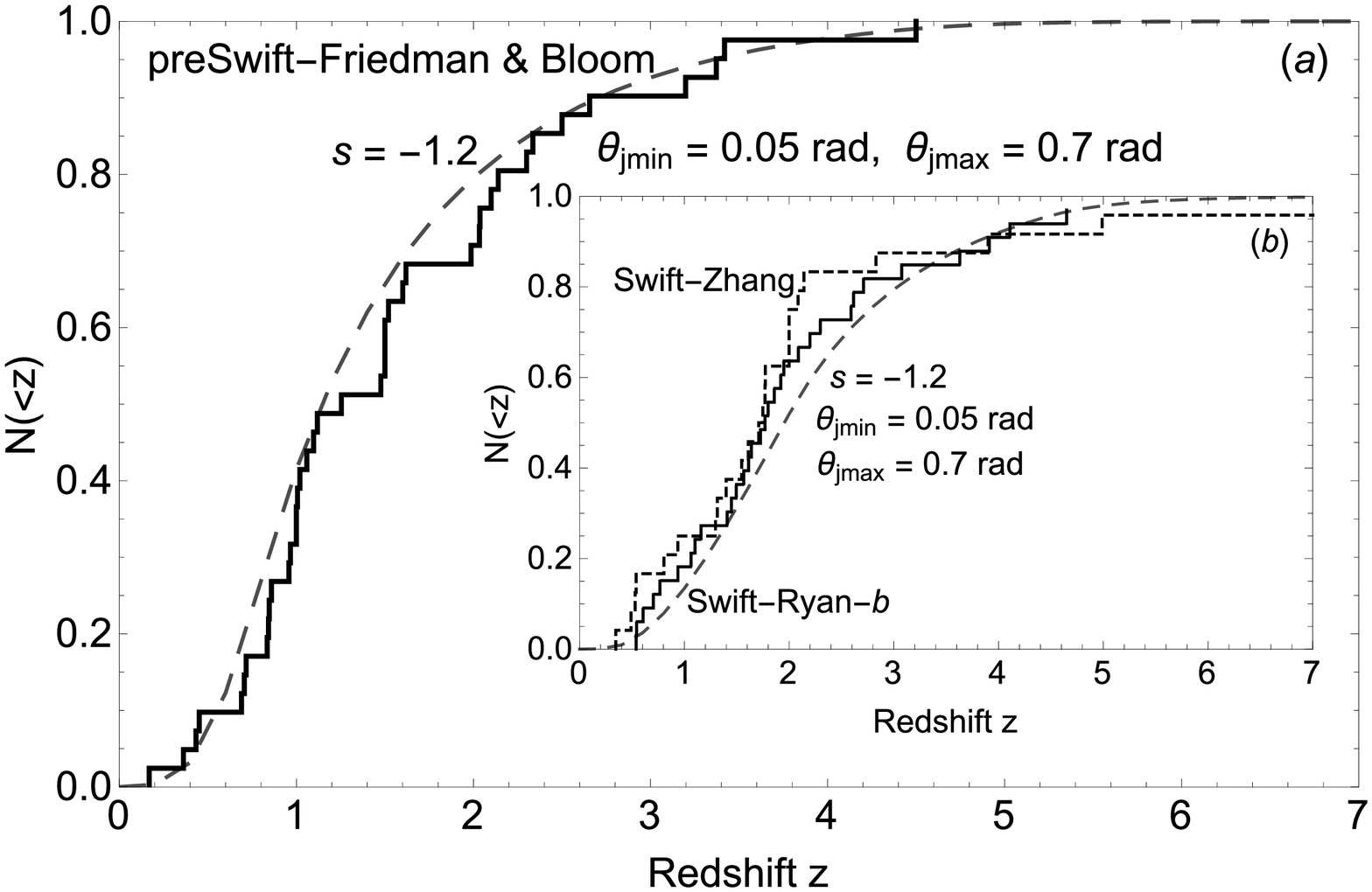}{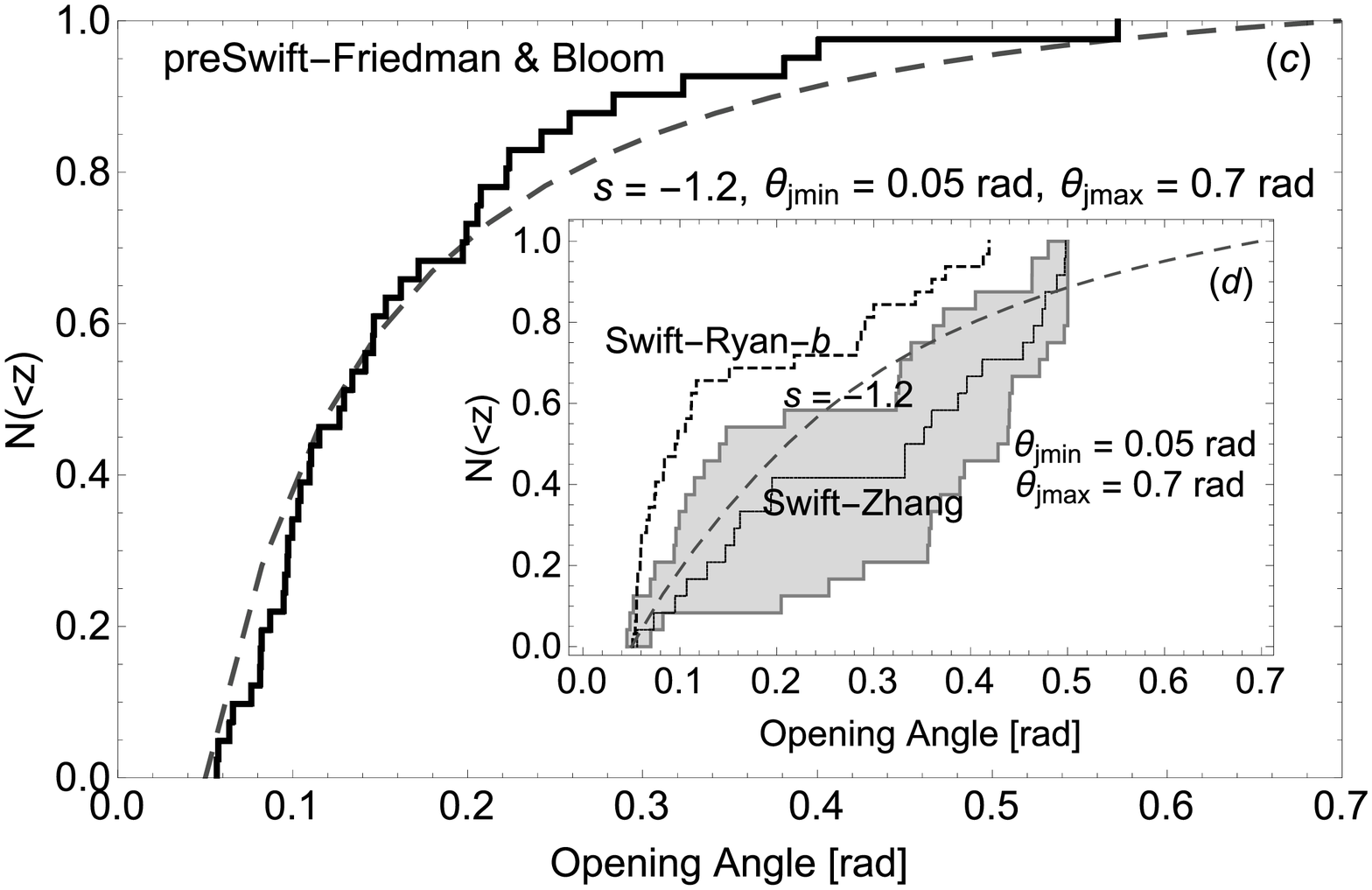}
\caption{\footnotesize (a) The cumulative redshift distributions of the fitted model (thin dashed line; also in (b)), the pre-\Swift~sample~(thick solid line), and (b) the \Swift-Zhang~(thick dashed line) and the \Swift-Ryan-b~(thick solid line) samples. (c) The cumulative jet opening angle distributions of the fitted model (thin dashed line; also in (d)), the pre-\Swift~(thick solid line), and (d) the \Swift-Ryan-b (thick dashed line) and the  \Swift-Zhang (thick solid line) samples with error bars (shaded region). These results use $\theta_{\rm j,min} = 0.05$ rad, $\theta_{\rm j,max} = 0.7$ rad, $\Estarg = 2 \times 10^{51}$ erg, and $s=-1.2$ with SFR9.}
\label{fig5} 
\finfig
\begfig[t] \hskip-0.25in \epsscale{1.15} \plottwo{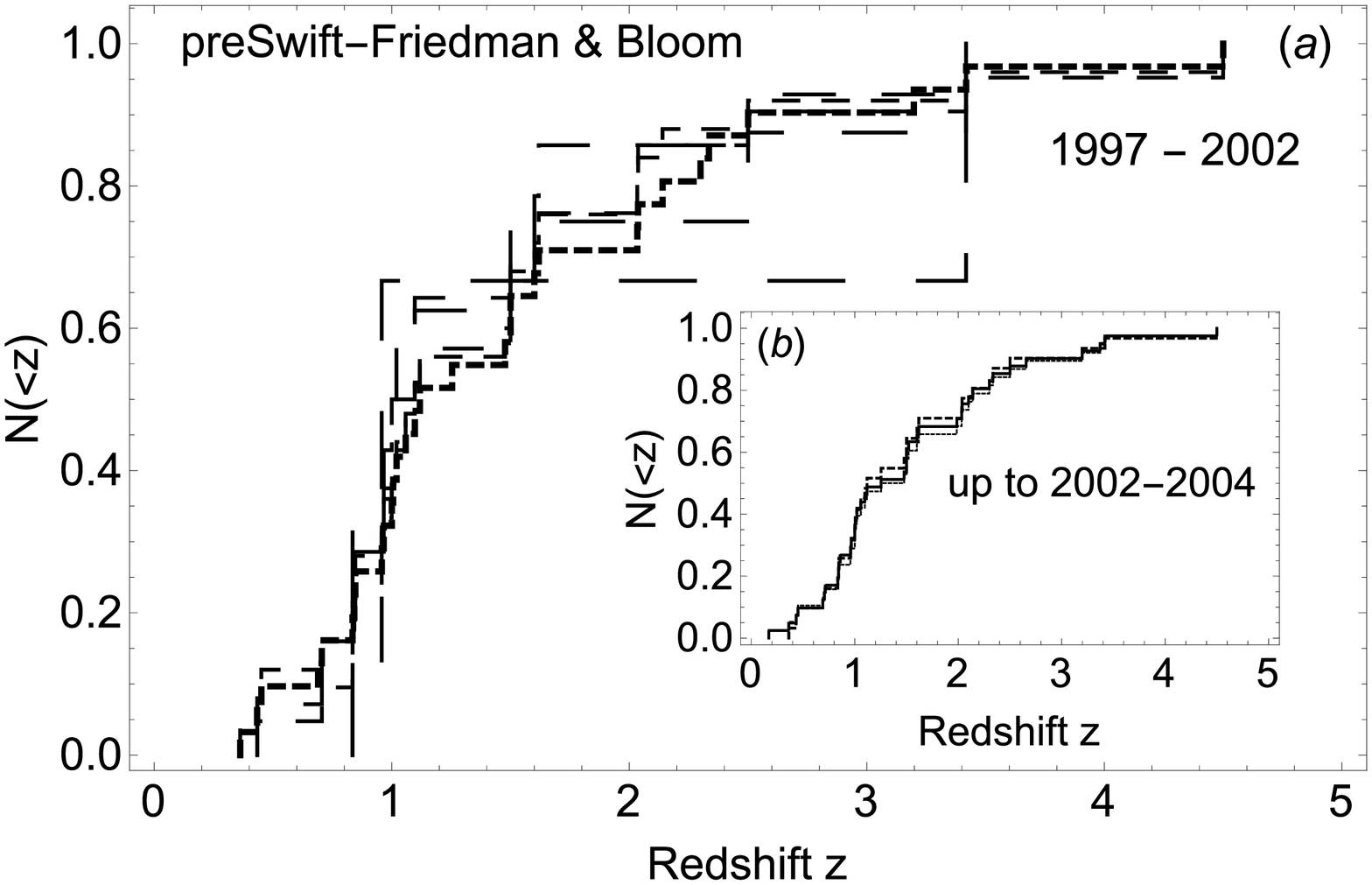}{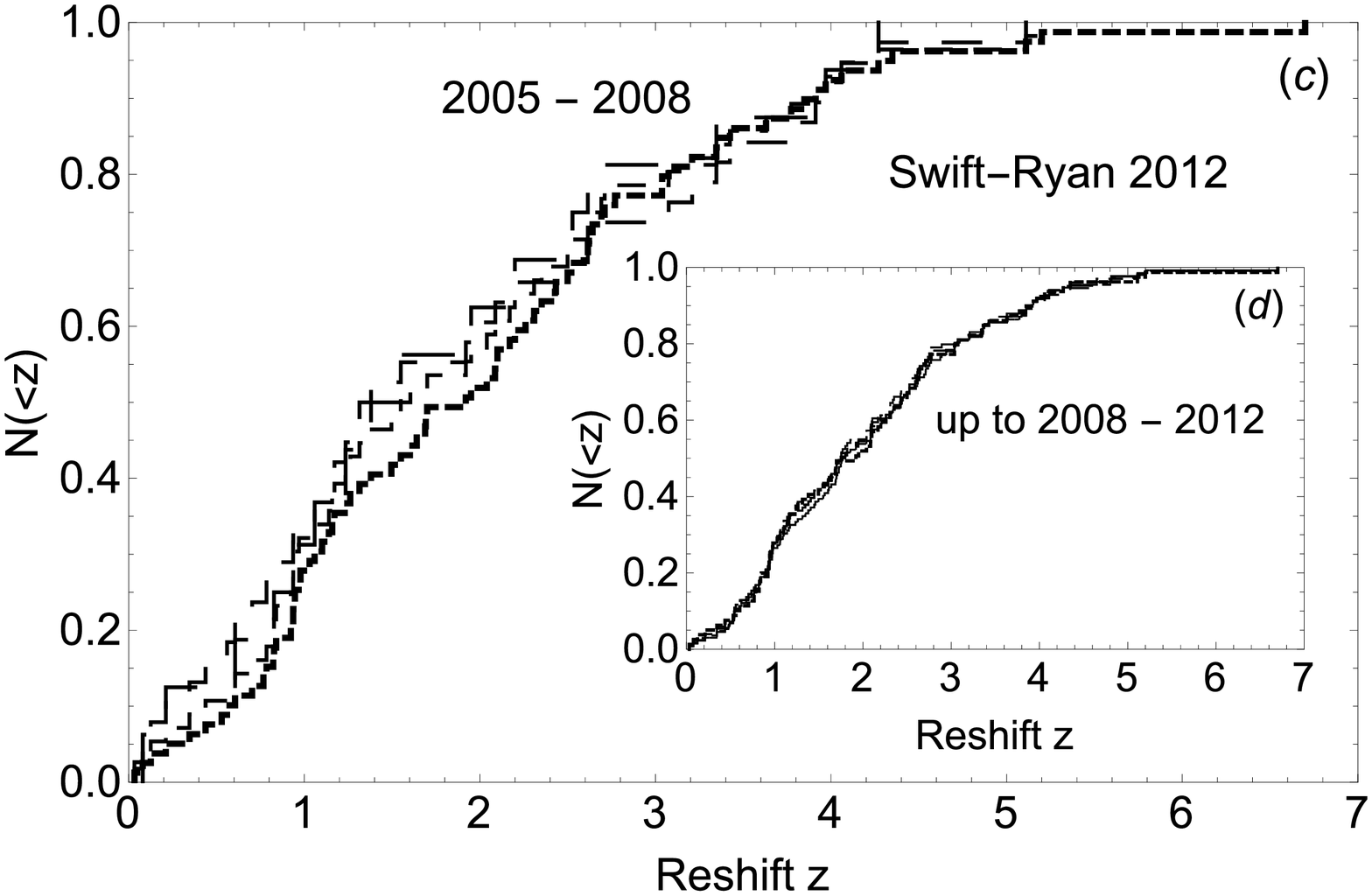}
\caption{\footnotesize Incomplete (a) pre-\Swift~and (c) \Swift~redshift samples from 1997 up to 2002 and from 2005 up to 2008, respectively. Complete (b) pre-\Swift~and (d) \Swift~redshift samples up to 2002 through 2004 and up to 2008 through 2012, respectively. The dark dashed lines in the panels (a) and (c) are data up to 2002 and 2008, respectively.}
\label{fig6} 
\finfig
\begfig[t] \hskip-0.25in \epsscale{1.15} \plottwo{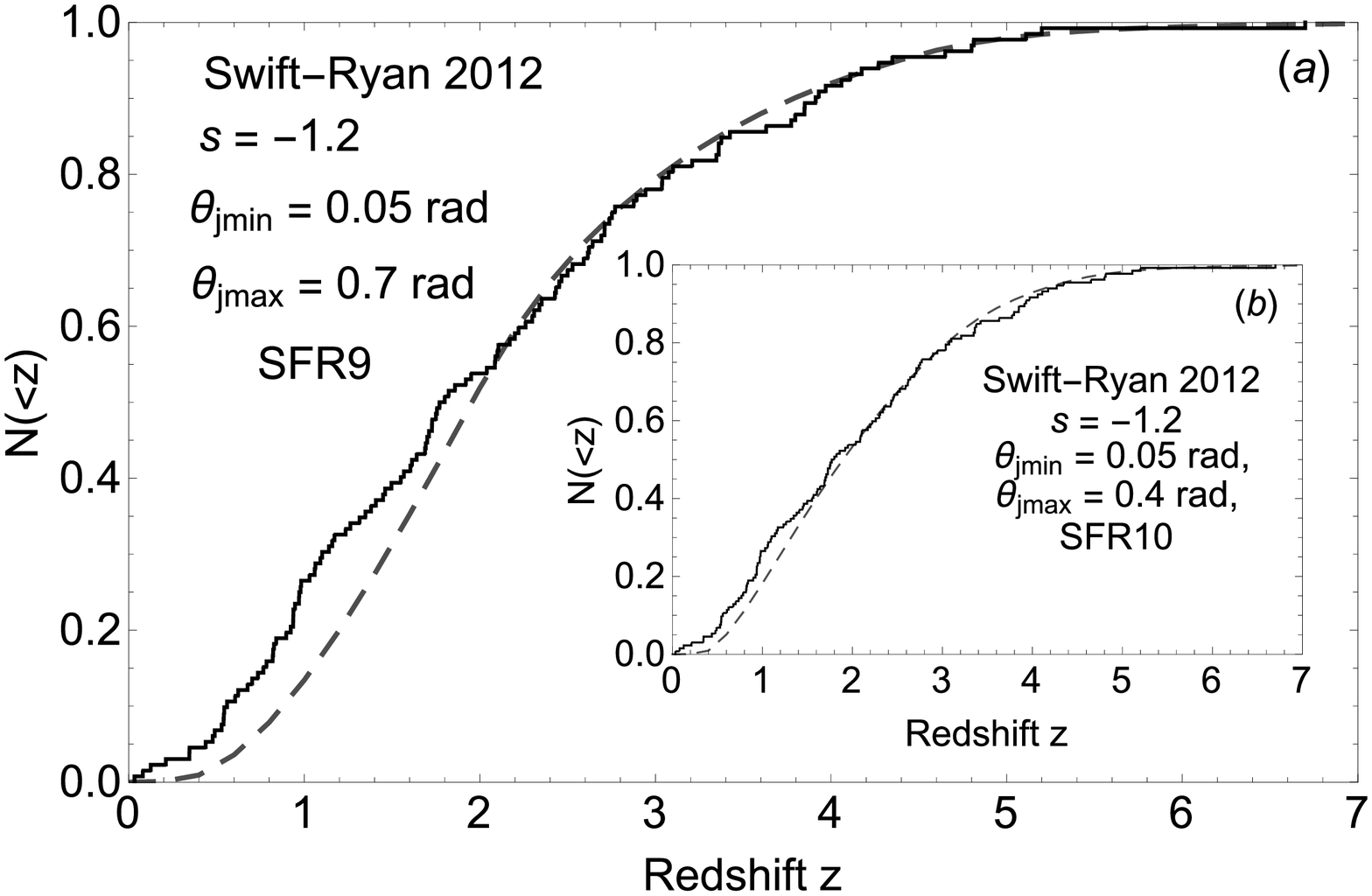}{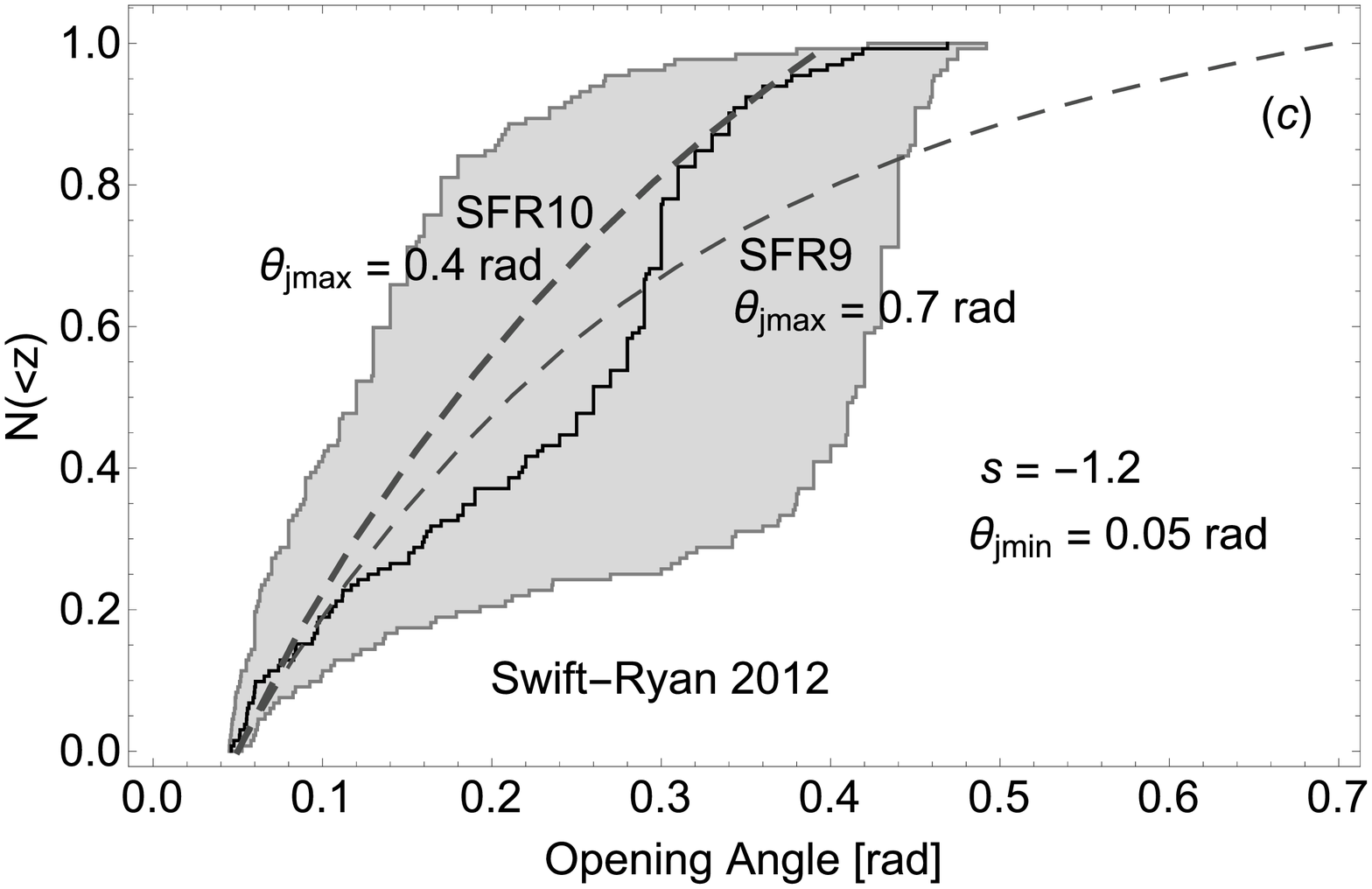}
\caption{\footnotesize (a) The cumulative redshift distributions of the fitted model (dashed line) using SFR9 and (b) the fitted model (dashed line) using SFR10 with the \Swift-Ryan 2012 (solid line) sample. (c) The cumulative jet opening angle distributions of the fitted model using SFR9 (thin dashed line) and SFR10 (thick dashed line) with the \Swift-Ryan 2012 sample~(solid line) and error bars (shaded region). These results use $\Estarg = 2 \times 10^{51}$ erg, $\theta_{\rm j,min} = 0.05$ rad, and SFR9 and SFR10 with $\theta_{\rm j,max} = 0.7$ and $0.4$ rad, respectively.}
\label{fig7} 
\finfig
\begfig[t] \hskip-0.25in \epsscale{1.15} \plottwo{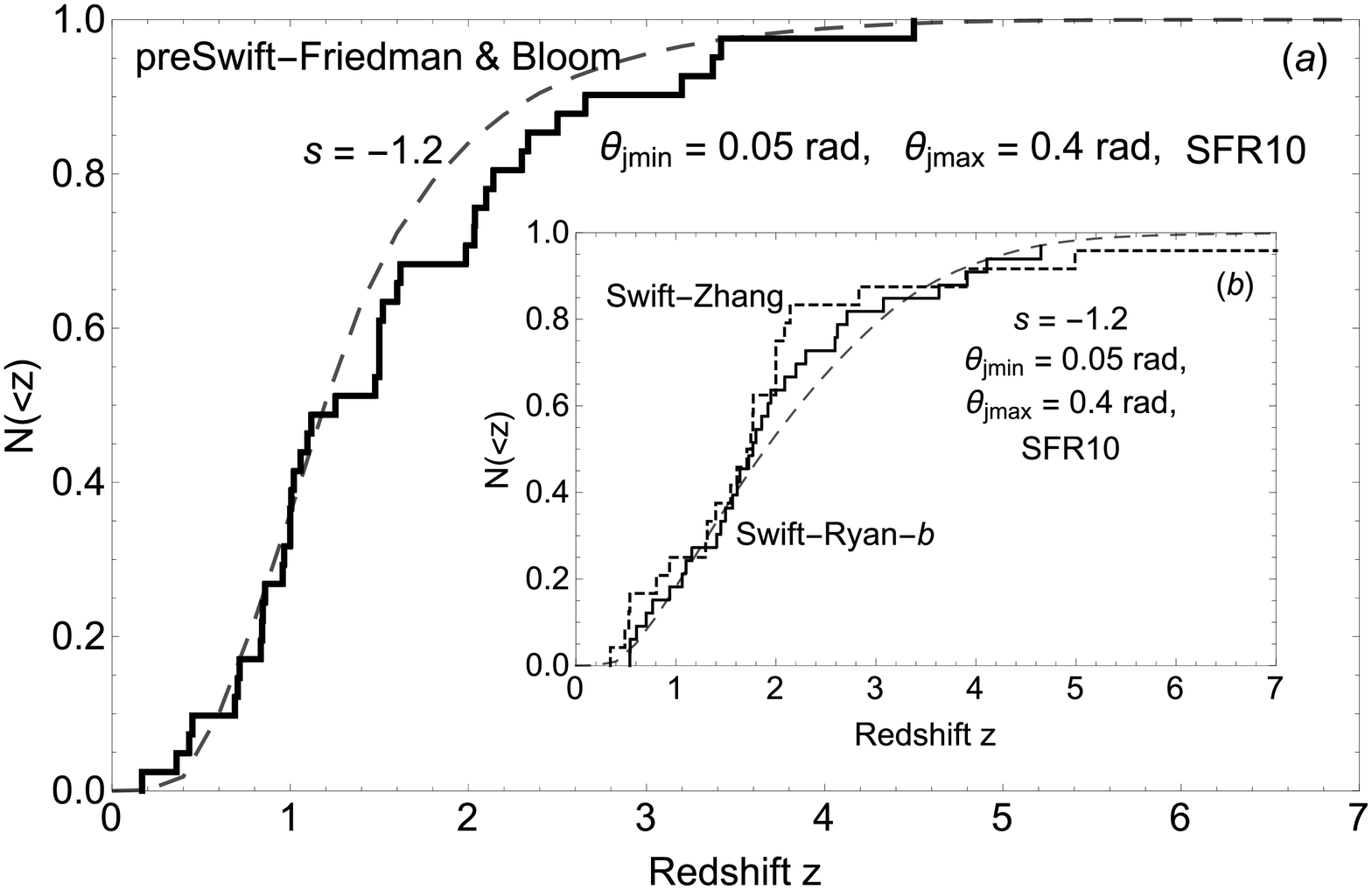}{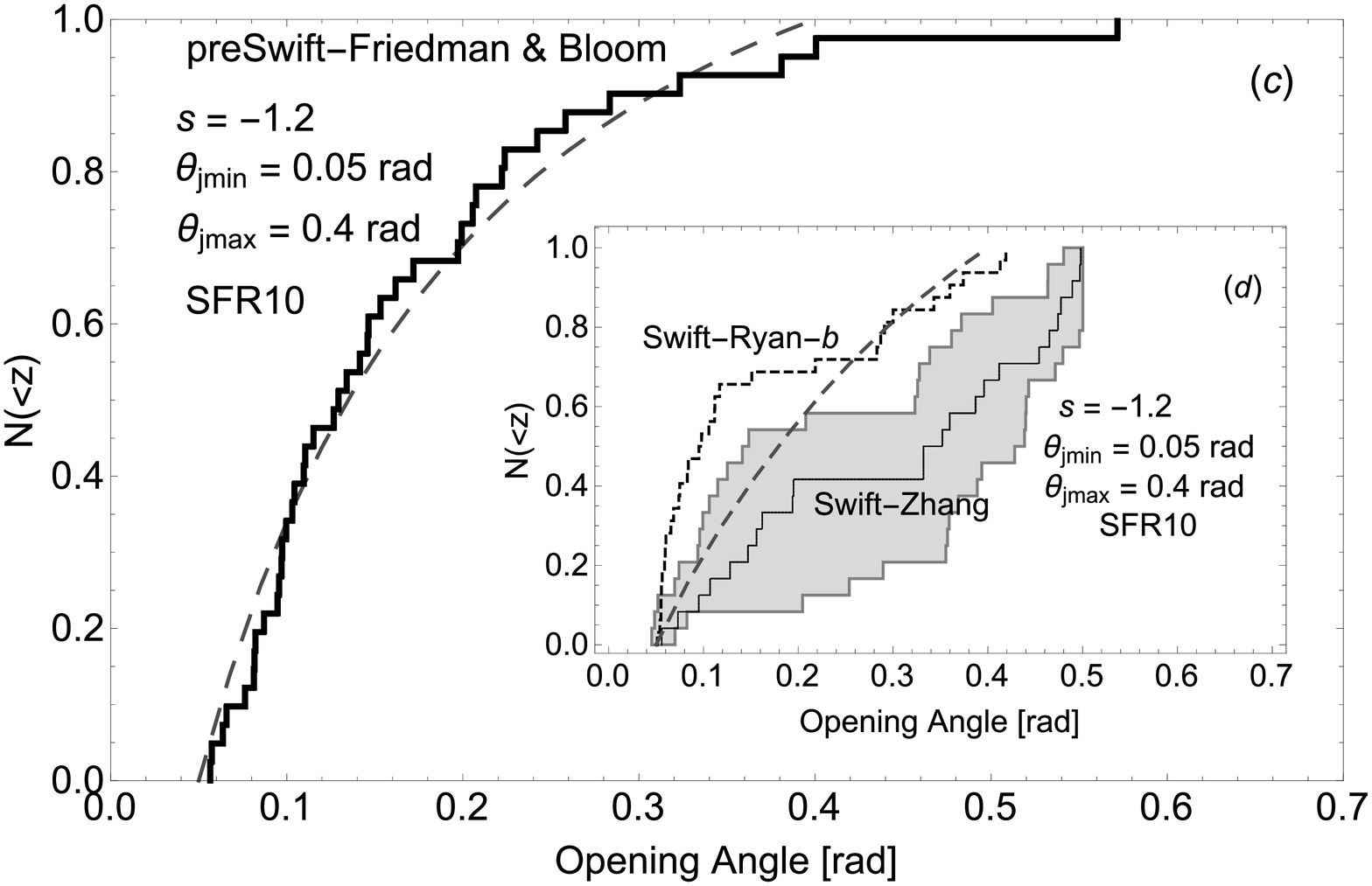}
\caption{\footnotesize (a) The cumulative redshift distributions of the fitted model (thin dashed line), the pre-\Swift~(thick solid line), and (b) the \Swift-Zhang (thick dashed line) and \Swift-Ryan-b~(thick solid line)~samples. (c) The cumulative jet opening angle distributions of the fitted model (thin dashed line), the pre-\Swift~sample~(thick solid line), and (d) the \Swift-Zhang (solid line) with error bars (shaded region) and \Swift-Ryan-b (thick dashed line) samples. These results use $\theta_{\rm j,min} = 0.05$ rad, $\theta_{\rm j,max} = 0.4$ rad, $\Estarg = 2 \times 10^{51}$ erg, and $s=-1.2$ with SFR10.}
\label{fig8} 
\finfig
\begfig[t] \hskip-0.25in \epsscale{0.8} \plotone{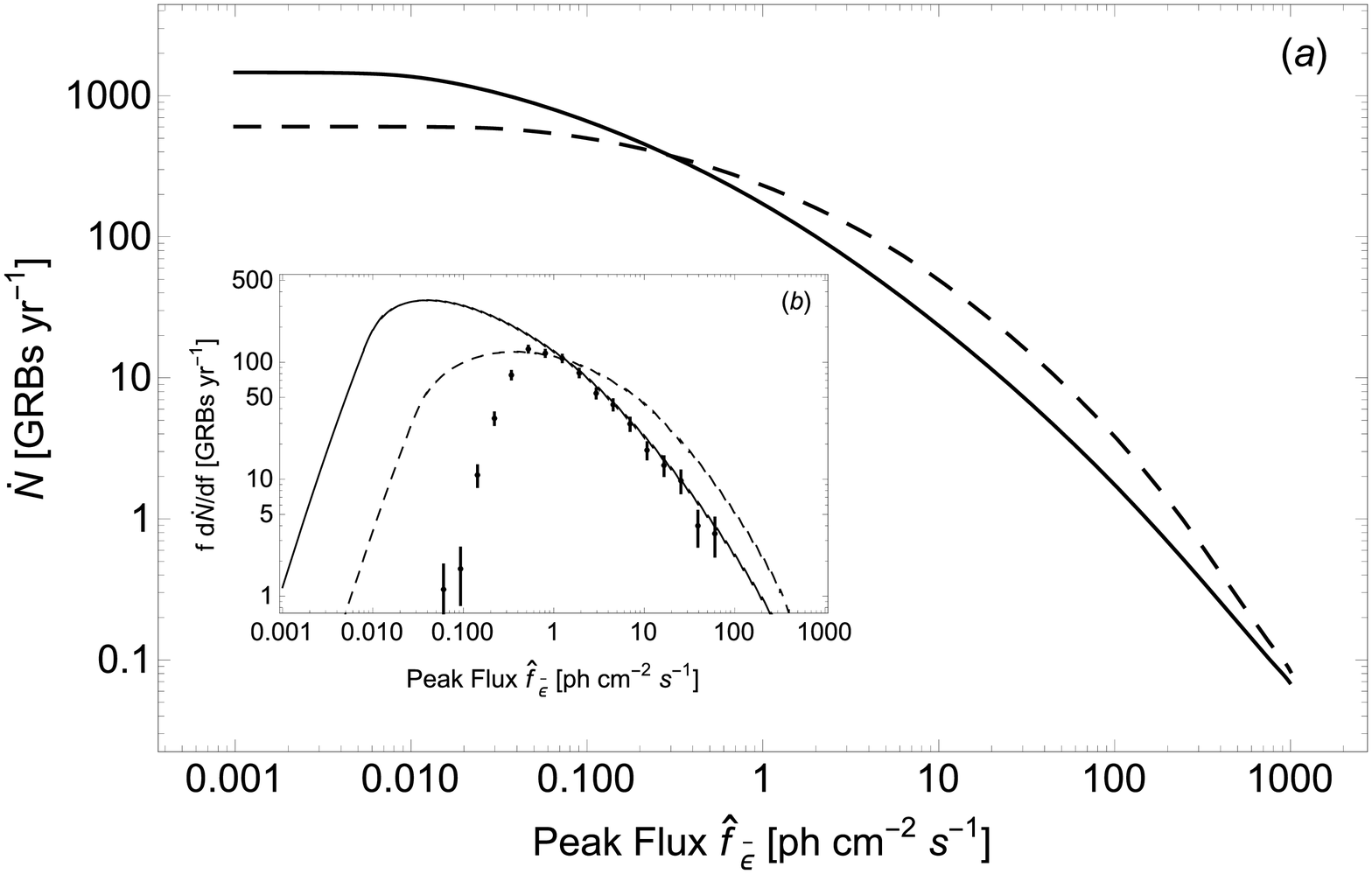}
\caption{\footnotesize Integral size distributions (a) and the differential size distributions (b) for SFR9 (solid line) and SFR10 (dashed line). The filled circle curve represents the $1024$ ms trigger timescale from the 4B catalog data, containing 1292 bursts that normalize to 371 LGRBs above a photon number threshold of $0.0633 \ \rm ph \ cm^{-2} \ s^{-1}$, and where the error bars are the statistical error in each bin.}
\label{fig9} 
\finfig

Nevertheless, it is clear that the above results are subject to selection bias because we fit only to a subset of the \Swift~GRBs  redshift sample. It is, therefore, interesting to see if our model can fit the complete \Swift-Ryan-2012 redshift and jet opening angle samples.  To determine if the \Swift-Ryan-2012 \citep[see Table 4 in][]{rya15} and pre-\Swift~\citep[see Table 2 in][]{fb05} sampling sizes are complete, we sort the pre-\Swift~and \Swift~data by year. The curves in Figure~\ref{fig6}(a) represent cumulative data in cumulative years, for example, data in 1997, 1997-1998, 1997-1999, and so on up to 2002, and then 1997-2002 up to 2004 in Figure~\ref{fig6}(b). We do the same for the \Swift-Ryan-2012 sample, as depicted in Figures~\ref{fig6}(c) and \ref{fig6}(d) for the years between 2005 and 2012. The results in Figures~\ref{fig6}(b) and (d) indicate that there is little variation to the redshift distribution after 2002 and 2008 for the pre-\Swift~and \Swift~samples, respectively, suggesting that the data are complete and any data that are collected after 2002 and 2008 for pre-\Swift~and \Swift, respectively, are not necessary. However, from Figures~\ref{fig6}a and (b), it is hard to say in general that the pre-\Swift~redshift data are complete since there are more than 57 GRBs that were detected by INTEGRAL after 2007 with unknown redshift. 

Moving toward a more complete redshift \Swift~sample, we search for a GRB burst rate functional form that could fit only the \Swift-Ryan-2012 redshift while predicting (1) the jet opening angle distribution for the \Swift-Ryan-2012 sample and (2) the redshift and jet opening angle distributions for the pre-\Swift~sample. Interestingly, the calculated redshift distribution using SFR9 fits well the observed estimated \Swift-Ryan-2012 redshift distribution above a redshift of 2 but is way off below it (see Figure~\ref{fig7}(a)), while the calculated jet opening angle distribution from our model is within the \Swift-Ryan-2012 statistical error bars (see Figure~\ref{fig7}(c)). The result of the fit to the redshift distribution suggests that we have an excess of GRBs at low redshift using SFR9. This means to fit the \Swift-Ryan-2012 redshift distribution we have to increase the GRB rate at lower redshift, and by doing so we shift the distribution to lower redshift, and consequently, we also shift the distribution more to the left at higher $z$. To finally fit the \Swift-Ryan-2012 redshift distribution, we have to decrease the maximum jet opening angle to $\theta_{\rm j,max} = 0.4$ rad, which is still within the acceptable range for LGRBs. The best-fit parameters have the range of jet opening angles between $\theta_{\rm j,min} = 0.05$ rad and $\theta_{\rm j,max} = 0.4$ rad, the jet opening angle power-law index $s = -1.2$, and the $\gamma$-ray energy released $\Estarg = 2 \times 10^{51}$ erg. These results are shown in Figures~\ref{fig7}(b) and \ref{fig7}(c), and the fits indicate an acceptable fit to the \Swift-Ryan 2012 redshift and jet opening angle distributions using SFR10 (using Equation~(\ref{eq18}) with $\alpha = 8, \beta = -0.4, \gamma = -5.1, z_1 = 0.5, z_2 = 4.5$; see Figure~\ref{fig1}(b)). Using SFR10 we also obtain acceptable fits to the pre-\Swift~sample, as shown in Figures~\ref{fig8}(a) and \ref{fig8}(c). The fitted parameters also give acceptable fits to the \Swift-Zhang and \Swift-Ryan-b redshift distributions and are within the estimated jet opening angle distributions between the \Swift-Zhang and \Swift-Ryan-b samples. For the first time, we have shown self-consistently that the GRB formation rate does follow the observed SFR, similar to the \citet{hb06} star formation history and as extended by Li (2008; SFR7), consistent with, for example, the \citet{wp10}, \citet{dd14}, \citet{gre15}, \citet{szl15}, and \citet{per16} analyses. 

The SFR10 functional form, however, indicates that the GRB burst rate is higher than the observed SFR7 at all redshifts, suggesting the possibility that the burst rate could be affected by luminosity evolution with redshift~\citep[e.g.,][]{sc07,sal09,sal12,den16,per16}.  On the other hand, the SFR9 functional form indicates that, between redshift 0.5 and 3, the GRB burst rate is lower than the observed SFR7, indicating an excess of GRBs at lower redshifts (see Figure~\ref{fig7}(a)) and therefore suggesting the possibility that the burst rate could be affected by different progenitor populations or GRB properties, for example, due to the increasing metallicities with decreasing redshift~\citep[e.g.,][]{fru06,ber07}. Interestingly, this excess of GRBs at low redshift is also consistent with, for example,  the \citet{yu15} and \citet{pet15} findings.  However, an interesting point to note is that in our GRB model we use a single-average value of gamma-ray energy released, $\Estarg = 2 \times 10^{51}$ erg, to represent the gamma-ray energy released by all LGRBs. Since the observed estimated energy release in the GRB explosions is between $10^{48} - 10^{52}$ erg \citep[e.g.,][]{fb05,kz15}, if we relax our gamma-ray energy released restriction by considering a distribution of $\Estarg$ over redshift or jet opening angle \citep[e.g.,][]{kb08,cen10,kz15} that also includes the luminosity functions that permit a potentially large number of low-luminosity events, where GRBs at higher redshift are brighter than those at lower redshift~\citep[e.g.,][]{sal12,den16}, then this might improve our fits to the pre-\Swift~and \Swift~redshift and jet opening angle distributions while reducing the GRB formation density rate (SFR10 case) over all redshifts or increasing the GRB formation density rate (SFR9 case) at low redshift to match the observed star formation density rate that is similar to the SFR7 from \citet{hb06} and as extended by \citet{li08}. We plan to explore this idea in a future paper. 

More importantly, in Figures~\ref{fig5}(d) and \ref{fig7}(c), our model predictions for the \Swift~jet opening angle distributions are within the error bars estimated by \citet{zha15} and \citet{rya15} from the subset and complete \Swift~samples, respectively. However, the median jet opening angles obtained by \citet{zha15} and \citet{rya15} are anywhere in the range $\langle \theta_{\rm j} \rangle \sim [0.1 - 0.35]$ rad, indicating problems with estimating the jet opening angles from the \Swift~data by combining the late-time {\it Chandra} data with well-sampled \Swift/XRT light curve observations \citep[e.g.,][]{lia08}. In fact, \citet{wan15} have systematically looked into the consistency between optical and X-ray afterglows, and they find that the final sample that consists of the same jet break with achromatic break times is small for the ``bright'' sample of GRBs. 

\subsection{GRB Count Rates and Size Distributions for \Swift~and BATSE}

Our model parameters suggest that GRBs can be detected with \Swift to a maximum redshift $z \approx 18$. The models, SFR9 and SFR10, that fit the data imply that about 10\% of Swift GRBs occur at $z \gtrsim 4$, in accord with the data as shown in Figures~\ref{fig7}(a) and \ref{fig7}(b). Our model also predicts that about 5\% of GRBs should be detected from $z\geq 5$. If more than 5\% of GRBs with $z > 5$ are detected by the time $\approx 100$ \Swift~GRBs with measured redshifts are found, then we may conclude that the GRB formation rate power-law index $\gamma$ (see Equation~\ref{eq18}) is shallower than our fitted value $\gamma = -5.1$ above $z\approx 4.5$. By contrast, if a few $z\gtrsim 5$ GRBs have been detected, then this would be in accord with our model and would support the conjecture that the LGRB burst rate follows the SFR that is similar to the \citet{hb06} and as extended by \citet{li08} SFR history. Moreover, the models also indicate that  less than 10\% (20\%) of LGRBs should be detected at $z \lesssim 1$ for SFR9 (SFR10), respectively. After examining the GRB samples\footnote{Redshifts were obtained from the catalogs maintained by Dan Perley http://www.astro.caltech.edu/grbox/grbox.php and Jochen Greiner http://www.mpe.mpg.de/$\sim$jcg/grbgen.html.} between 2013 and 2015, we notice that less than $5\%$ of LGRBs per year were detected by \Swift~above $z > 5$, and about $\sim 10\%$ of LGRBs were detected above $z > 4$, consistent with both model predictions. However, about 6\% of LGRBs per year were detected below $z < 1$, consistent more with model SFR9.  

We also compare our model size distribution with the BATSE 4B Catalog size distribution. The BATSE 4B Catalog contains 1292 bursts in total, including short- and long-duration GRBs. There are a total of 872 LGRBs that are identifiable in the BATSE 4B Catalog \citep{pac99,ld09} with a BATSE detection rate of $\approx 550$ GRBs per year over the full sky brighter than 0.3 photons $\rm cm^{-2} \, s^{-1}$ in the $50-300$ keV band \citep{ban02}. Since there are only 872 LGRBs in the catalog, the BATSE detection rate is $\approx 371$ LGRBs per year over the full sky. Figure \ref{fig9}(a) shows the model integral size distribution (see Equation~(\ref{eq9})) of LGRBs predicted by our best-fit models, SFR9 and SFR10. The plots are normalized to the current total number of observed LGRBs per year from BATSE, which is 371 bursts per $4 \pi$ sr exceeding a peak flux of 0.3 photons $\rm cm^{-2} \ s^{-1}$  in the $\Delta E=50-300$ keV band for the $\Delta t=1.024$ s trigger time. From the model size distribution, we find that $\cong 235 (270)$ LGRBs per year should be detected with a BATSE-type detector over the full sky above an energy flux threshold of $\sim 10^{-7} \ \rm erg \ cm^{-2} \ s^{-1}$, or photon number threshold of $> 0.633 \ \rm ph \ cm^{-2} \ s^{-1}$ using SFR9 (SFR10), respectively. We also estimate from our fits that $\cong 800 (530)$ LGRBs take place per year per $4 \pi$ sr with a flux $\gtrsim 10^{-8} \ \rm erg \ cm^{-2} \ s^{-1}$, or $> 0.0633 \ \rm ph \ cm^{-2} \ s^{-1}$. The field of view of the BAT instrument on \Swift~is 1.4 sr \citep{geh04}, implying that \Swift should detect $\cong 90 (60)$ LGRBs per year for SFR9 (SFR10), respectively. Currently, \Swift~observes about 135 GRBs per year, and applying the ratio 872/1292, we obtain $\cong 90$ LGRBs per year; and this is consistent with the SFR9 model prediction. 

Finally, we show the differential size distribution of BATSE GRBs from the Fourth BATSE catalog \citep{pac99} in comparison with our model prediction in Figure~\ref{fig9}(b). As can be seen, our model parameters, using SFR9, give an excellent fit to the BATSE LGRB size distribution, while SFR10 overpredicts the differential burst rate above $1.0 \ \rm ph \ cm^{-2} \ s^{-1}$. It is important to note that the size distribution of the BATSE RGB distribution in Figure~\ref{fig9}(b) is normalized and corrected to 371 LGRBs above a photon number threshold of $0.3 \ \rm ph \ cm^{-2} \ s^{-1}$. Below a photon number threshold of $0.3 \ \rm ph \ cm^{-2} \ s^{-1}$ in the 50 -- 300 keV band, the observed number of GRBs falls rapidly due to the sharp decline in the BATSE trigger efficiency at these photon fluxes \citep[see][]{pac99}, while the size distribution of the \Swift~GRBs will extend to much lower values $\cong 0.0633 \ \rm ph \ cm^{-2} \ s^{-1}$. 

\section{Conclusions}
The answer to the question whether LGRBs follow the SFR is perhaps within sight. \citet{ld07} had examined whether the differences between the pre-\Swift~and \Swift~redshift distributions can be explained with a physical model for GRBs that takes into account the different flux thresholds of GRB detectors. The conclusion of their work suggested that GRB activity was greater in the past and is not simply proportional to the bulk of the star formation as traced by the blue and UV luminosity density of the universe. The \citet{ld07} result for SFR density history at high redshift was consistent with that of \citet{mes06}, \citet{dai06}, and \citet{gp07} and was later confirmed by many other researchers \citep[e.g.,][]{kis08,yuk08,kis09,wp10,vir11,jak12,wan13}. However, it is unclear whether this excess at high redshift is due to some sort of evolution in an intrinsic luminosity \citep[e.g.,][]{sc07,sal09,sal12,den16,per16} or the cosmic evolution of the GRB rate \citep[e.g.,][]{but10,qin10,wp10}. Furthermore, \citet{sal12}, for example, suggested that a broken power law luminosity evolution with redshift is required to fit the observed redshift distribution, while other researchers proposed that it is not necessary to invoke luminosity evolution with redshift to explain the observed GRB rate at high $z$, by carefully taking selection effects into account \citep[e.g.,][]{wan13,den16}, or that GRBs do simply follow the SFR \citep[e.g.,][]{wp10,dd14,gre15,szl15,per16}. Since it is unclear whether the excess of GRB rate at high redshift is due to luminosity evolution or some other means, or that GRB rates do follow the SFR, we revisit the work done by \citet{ld07} by moving away from the GRB energy flat spectrum and using a more complete GRB data sample from pre-\Swift~and \Swift~instruments. 

Considering a broken power law $\nu F_\nu$ GRB SED with $a =1$ and  $b = -0.5$ as the low- and high-energy indices, respectively, the flux thresholds $\fluxthres$ are set equal to $10^{-8}$ erg cm$^{-2}$ s$^{-1}$ and $10^{-7}$ erg cm$^{-2}$ s$^{-1}$ for \Swift~and pre-\Swift, respectively. Assuming that the GRB properties do not change with time, we find that our GRB model provides acceptable fits to the complete pre-\Swift~and \Swift~redshift and jet opening angle distributions with the best-fit parameters having the range of jet opening angles between $\theta_{\rm j,min} = 0.05$ rad and $\theta_{\rm j,max} = 0.4$ rad, the jet opening angle power law index $s = -1.2$, and the $\gamma$-ray energy released $\Estarg = 2 \times 10^{51}$ erg with the GRB source density rate (SFR10; using Equation~(\ref{eq18}) with $\alpha = 8, \beta = -0.4, \gamma = -5.1, z_1 = 0.5, z_2 = 4.5$) that follows the density star formation history similar to \citet{hb06} and extended by \citet{li08}. While using a subset of the \Swift~data, our model also gives acceptable fits to the redshift and jet opening angle distributions with the best-fit parameters having the range of jet opening angles between $\theta_{\rm j,min} = 0.05$ rad and $\theta_{\rm j,max} = 0.7$ rad, the jet opening angle power-law index $s = -1.2$, and the $\gamma$-ray energy released $\Estarg = 2 \times 10^{51}$ erg with the GRB source density rate (SFR9; using Equation~(\ref{eq18}) with $\alpha = 4.1, \beta = 0.8, \gamma = -5.1, z_1 = 0.5, z_2 = 4.5$) that is also similar to SFR10. More importantly, the result, using SFR9, suggests that there is an excess of GRBs below a redshift of 2 when we apply this model to fit the complete \Swift~redshift and jet opening angle distributions. This result could perhaps indicate that different properties of GRBs are occurring at lower redshift (e.g., high-metallicity GRBs) than the higher redshift (e.g., low-metallicity GRBs) counterpart \citep[e.g.,][]{fru06,ber07}. We plan to explore these behaviors in a future paper by examining GRBs that have similar spectra properties. 

We checked our model prediction for the number of GRBs that can be detected with \Swift~and show that about 10\% (5\%) of Swift GRBs occur at $z \gtrsim 4 (5)$, respectively, for both SFR9 and SFR10. Moreover, the models also indicate that  less than 10\% (20\%) of LGRBs should be detected at $z \lesssim 1$ for SFR9 (SFR10), respectively. After examining the GRB samples, we notice that less than $5\%$ of LGRBs per year were detected by \Swift~above $z > 5$, and about $\sim 10\%$ of LGRBs were detected above $z > 4$, consistent with both model predictions. However, about 6\% of LGRBs per year were detected below $z < 1$, in favor of model prediction SFR9.  Second, we examine our model size distribution with the BATSE 4B Catalog size distribution, which contains 1292 bursts total, with 872 being LGRBs, and we find that $\cong 235 (270)$ LGRBs per year should be detected with a BATSE-type detector over the full sky above an energy flux threshold of $\sim 10^{-7} \ \rm erg \ cm^{-2} \ s^{-1}$, or photon number threshold of $> 0.633 \ \rm ph \ cm^{-2} \ s^{-1}$ using SFR9 (SFR10), respectively, while $\cong 800 (530)$ LGRBs take place per year per $4 \pi$ sr with a flux $\gtrsim 10^{-8} \ \rm erg \ cm^{-2} \ s^{-1}$, or $> 0.0633 \ \rm ph \ cm^{-2} \ s^{-1}$. Since the field of view of the BAT instrument on \Swift~is 1.4 sr \citep{geh04}, this implies that Swift should detect $\cong 90 (60)$ LGRBs per year for SFR9 (SFR10), respectively. Currently, \Swift~observes about 135 GRBs per year, and applying the ratio 872/1292, we obtain $\cong 90$ LGRB per year; this is also consistent with the SFR9 model prediction. Finally, we show the differential size distribution of BATSE GRBs from the Fourth BATSE catalog \citep{pac99} in comparison with our model prediction in Figure~\ref{fig9}(b). Using our model parameters, SFR9 also gives an excellent fit to the BATSE LGRB size distribution, while SFR10 overpredicts the differential burst rate above $1.0 \ \rm ph \ cm^{-2} \ s^{-1}$. 

In conclusion, it is clear that our analysis leans toward GRB formation rate model SFR9, and that this GRB formation rate follows a star formation history similar to \citet{hb06} and extended by \citet{li08}. More importantly, SFR9 provides (1) a self-consistent prediction for the total number of GRBs per year at $z > 4$, $ z < 1$, and over the full sky for \Swift, and (2) an excellent fit to the BATSE LGRB size distribution. Furthermore, this model indicates that there is an excess of GRBs at low redshift, consistent with the findings of other researchers; it also predicts the jet opening angle distribution that lies within the \Swift~observed estimated jet opening angle distribution error bars. However, the \Swift~observed estimated jet opening angles currently contain large error bars, and perhaps, with a better estimated jet opening angle distribution and the future fraction of low- and high-$z$ standard long-duration GRBs from \Swift, this model can be tested.

\acknowledgements T.L. wishes to acknowledge C. Dermer, B. Zhang, B.B. Zhang, A. Friedman, and N. Gehrels for discussions and correspondence and the anonymous referee for comments and useful suggestions that improved the paper.

\end{document}